\documentclass[12pt,fleqn]{article} 
\usepackage{graphicx}
\renewcommand{\Im}{\mbox{Im }}
\renewcommand{\Re}{\mbox{Re }}
\newcommand{\ra}{\rightarrow}
\newcommand{\bra}{\langle} \newcommand{\ket}{\rangle}
\newcommand{\be}{\begin{equation}}
\newcommand{\ee}{\end{equation}}
\newcommand{\bea}{\begin{eqnarray}}
\newcommand{\eea}{\end{eqnarray}}
\newcommand{\eps}{\epsilon}

\newcommand{\ffi}{\varphi}

\newcommand{\ep}{\qquad {\vrule height 10pt width 8pt depth 0pt}}

\newcommand{\grintl}{[\kern-.18em [}
\newcommand{\grintr}{]\kern-.18em ]}
\newcommand{\ds}{\displaystyle}
\newtheorem{lem}{Lemma}[section]
\newtheorem{prop}{Proposition}[section]
\newtheorem{thm}{Theorem}[section]
\newtheorem{cor}{Corollary}[section]

\def\smallR{\hbox{\scriptsize I\kern-.23em{R}}}
\def\R{\hbox{$\mit I$\kern-.33em$\mit R$}}
\def\C{\hbox{$\mit I$\kern-.6em$\mit C$}}
\def\un{\hbox{$\mit I$\kern-.77em$\mit I$}}
\def\0{\hbox{$\mit I$\kern-.70em$\mit O$}}
\def\r{I\kern-.277em R}

\def\N{\mbox{\bf N}}

\begin{document}

\title{Determination of Non--Adiabatic Scattering Wave Functions
in a Born--Oppenheimer Model}
\author{George A. Hagedorn\thanks{Partially
Supported by National Science Foundation
Grants DMS--0071692 and DMS--0303586.}\\
Department of Mathematics and\\
Center for Statistical Mechanics and Mathematical Physics\\
Virginia Polytechnic Institute and State University\\
Blacksburg, Virginia 24061-0123, U.S.A.\\[15pt]
\and
Alain Joye\\
Institut Fourier\\ Unit\'e Mixte de Recherche CNRS-UJF 5582\\
Universit\'e de Grenoble I\\
BP 74\\
F--38402 Saint Martin d'H\`eres Cedex, France}

\date{ }
\maketitle

\vskip 1.5cm
\begin{abstract}
We study non--adiabatic transitions in
scattering theory for the time dependent molecular Schr\"odinger
equation in the Born--Oppenheimer limit.
We assume the electron Hamiltonian has finitely many
levels and consider the propagation of coherent states 
with high enough total energy.

When two of the electronic levels are isolated from
the rest of the electron Hamiltonian's spectrum and
display an avoided crossing, we compute the component of the
nuclear wave function associated with the non--adiabatic
transition that is generated by propagation through the avoided crossing.
This component is shown to be exponentially small in the square of the
Born--Oppenheimer parameter, due to the Landau-Zener mechanism. It
propagates asymptotically as a free Gaussian in the nuclear
variables, and its momentum is shifted.
The total transition probability 
for this transition and the momentum shift are both larger
than what one would expect from a naive approximation and
energy conservation.
\end{abstract}

\newpage

\section{Introduction}

We study scattering theory for the time--dependent
molecular Schr\"odinger equation 
\be\label{schr}
i\,\eps^2\,\frac{\partial}{\partial t}\psi(x,t,\eps)\ =\ 
\left(\,-\,\frac{\eps^4}{2}\, 
\frac{\partial^2\phantom{x}}{\partial x^2}\ +\ 
h(x)\,\right)\,\psi(x,t,\eps)\qquad\mbox{in}\qquad
L^2(\R,\,\C^m), 
\ee 
where the electronic hamiltonian $h(x)$ is an $m\times m$
self-adjoint matrix that depends on the nuclear position
variable $x\in\R$. The Born-Oppenheimer parameter $\eps>0$
denotes the fourth root of the electron mass divided by
the mean nuclear mass. 

We compute the leading order asymptotics of nuclear wave
functions associated with certain non--adiabatic transitions
of the electrons. The Landau-Zener mechanism responsible for 
these makes them exponentially small
in $1/\eps^2$ as $\eps\ra 0$.

Our most general result can be found in Theorem \ref{mai}.
Describing the most general situation requires the development of
a significant amount of notation and some technical hypotheses. 
So, in this introduction, we describe two physically interesting
special cases that illustrate the main consequences 
of our analysis in a simple situation.
Theorems \ref{at_last_0} and \ref{at_last_m} give precise statements
of our results for these special cases.

Suppose $h(x)$ is a real $2\times 2$ self-adjoint matrix that depends
analytically on $x$ and has limits $h(\pm\infty)$ as $x\ra\pm\infty$
that are approached sufficiently rapidly.
Denote the eigenvalues of $h(x)$ by $e_j(x)$, and assume that
$e_2(x)\geq e_1(x)+\delta$ for all $x\in\R$, where $\delta>0$.
Near $x=0$, assume $e_1$ and $e_2$ have an avoided crossing,
{\it i.e.},  $e_2(x)-e_1(x)\simeq \sqrt{x^2+\delta^2}$ close 
to $x=0$, with $\delta$ small but positive. Such an avoided crossing 
corresponds to complex crossing points $z_0$ and $\overline{z_0}$,
where the analytic continuations of $e_1$ and $e_2$ satisfy
$e_1(z_0)=e_2(z_0)$, and $z_0$ is close to the real axis,
with $z_0=O(\delta)$. 

Let $\phi_1(x)$ and $\phi_2(x)$ denote normalized, real eigenvectors
associated with $e_1(x)$ and $e_2(x)$.

\vskip 8mm
\centerline{\includegraphics[height=2in,width=6in]{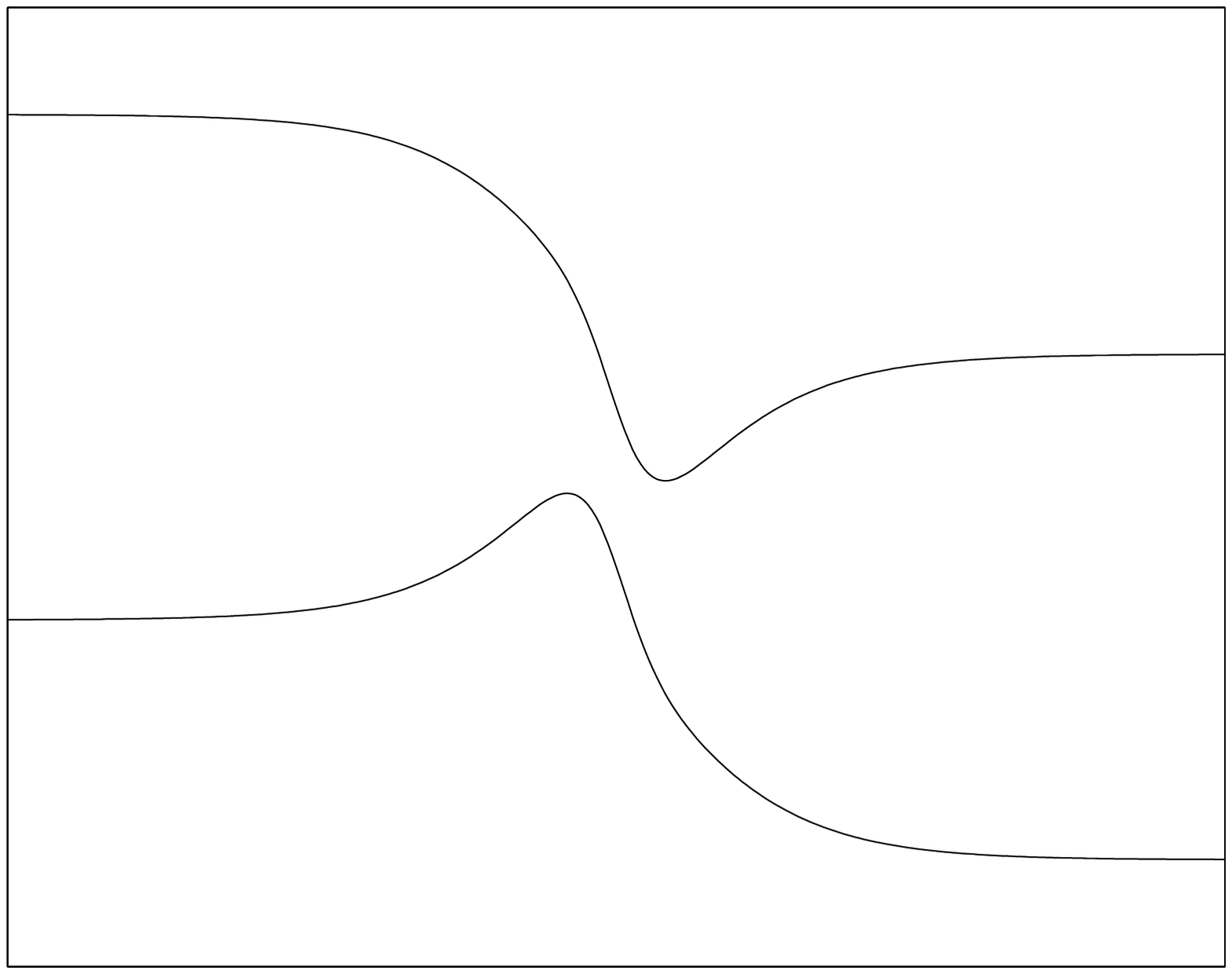}}

\vskip 5mm
\noindent {\bf Figure 1.} A plot of typical electron energy levels involved in
an avoided crossing. 

\vskip 8mm

Among the nuclear wave functions we can accommodate are Gaussian
coherent states that are defined by
\bea\nonumber\label{gcs}
\ffi_0(A,\,B,\,\eps^2,\,a,\,\eta,\,x)\ =\ 
\frac 1{\pi^{1/4}\,\eps^{1/2}\,A^{1/2}}\
\exp\left(\,-\,\frac{B\,(x-a)^2}{2\,A\,\eps^2}\,+\,
i\,\frac{\eta\,(x-a)}{\eps^2}\,\right),
\eea
where the complex numbers $A$ and $B$ satisfy the normalization
condition $\mbox{Re}\,\overline{B}A\,=\,1$.
These states are localized in position near $x=a$, and in momentum
near $p=\eta$. Their position uncertainty is $\eps|A|$ and their
momentum uncertainty is $\eps|B|$. For a thorough discussion of these
wave packets, see \cite{raise}.

Choose $E>\sup_{x\in\R}\,e_2(x)$. 
For a state incoming from the left on the upper electonic level,
choose $\eta_->0$. We assume $\eta_-$ is large enough so that the
classical
energy $\eta_-^2/2+e_2(-\infty)>E$. There exists a solution to (\ref{schr})
whose large negative $t$ asymptotics are given by 
\be\label{incoming0}
e^{i(\eta_-^2/2-e_2(-\infty))t/\eps^2}\,
\ffi_0(A_-+iB_-t,\,B_-,\,\eps^2,\,a_-+\eta_-t,\,\eta_-,\,x)\ \phi_2(x),
\ee
where the nuclear part is a free Gaussian.
Since the electronic levels are isolated from one another,
the large positive $t$ asymptotics of this solution are multiples of
$\phi_2(x)$, up to exponentially small errors in $1/\eps^2$. 
They have the leading behavior determined by the standard 
time--dependent Born--Oppenheimer approximation as $\eps\ra0$, see 
\cite{masterpiece}:
\bea\nonumber
e^{i\theta_1(\eps)}\,e^{i(\eta_1^2/2-e_2(\infty))t/\eps^2}\,
\ffi_0(A_1+iB_1t,\,B_1,\,\eps^2,\,a_1+\eta_1t,\,\eta_1,\,x)\ \phi_2(x),
\eea
where $e^{i\theta_1(\eps)}$ is some explicit phase, and the
parameters $A_1, B_1, a_1, \eta_1$ are determined by the 
scattering properties of the classical Hamiltonian $p^2/2+e_2(x)$.\\
 
Our interest lies with the leading order asymptotics of the
non--adiabatic component of the wave function for large
positive $t$ and $\eps\ra 0$. We prove in Theorem \ref{at_last_0}
that these have the form
\bea\nonumber
c_0\,e^{-\alpha^*/\eps^2}\,e^{i\theta_+(\eps)}\,
e^{i(\eta_+^2/2-e_1(\infty))t/\eps^2}\,
\ffi_0(A_++iB_+t,\,B_+,\,\eps^2,\,a_++\eta_+t,\,\eta_+,\,x)\ \phi_1(x),
\eea
and we specify how the phase $\theta_+(\eps)$, the $\eps$-independent 
amplitude $c_0>0$, the exponential decay rate $\alpha^*>0$, and the parameters of the
free Gaussian part $A_+$, $B_+$, $a_+$, and $\eta_+>0$ are
determined. As a corollary, the leading term of the transition
amplitude ${\cal A}(\eps)$ (whose absolute square is the transition
probability) 
is given by the quantity
\be\label{molLZ}
{\cal A}(\eps)=c_0\,e^{i\theta_+(\eps)}\, e^{-\alpha^*/\eps^2},
\ \ \ \mbox{as}\ \ \  \eps\ra 0.
\ee

Let us describe the main features of this exponentially small 
transmitted part of the wave function.
One may naively expect $\eta_+$ to be determined by the energy
conservation condition
\bea\nonumber
\frac{\eta_-^2}2\,+\,e_2(-\infty)\ =\ 
\frac{\eta_+^2}2\,+\,e_1(\infty),
\eea
but this yields the wrong value. The correct value is larger.
Intuitively, this is due to the faster parts of the wave function
behaving less adiabatically than the slower parts. Because this
dependence on the speed appears in an exponent, it leads to an
$O(1)$ change in the final momentum $\eta_+$.
In other words, the higher momentum components of the
incoming state are much more likely to make a transition
than the lower momentum components. Hence, after the transition,
there are more fast pieces of the wave function, and the 
final average mometum is greater than one would naively 
expect from an energy conservation
calulation based solely on the average incoming momentum.

This also affects the transition amplitude which is 
larger than what is naively expected. It is asymptotically 
composed of an $\eps$-independent prefactor $c_0$ times an exponentially
small quantity $e^{-\alpha^*/\eps^2}$, whose decay rate $\alpha^*$
is related to that of the Landau-Zener decay rate for purely 
adiabatic problems. Actually,  $\alpha^*$ consists of the sum
of the imaginary part of some action integral around the complex
electronic eigenvalue crossing point $z_0$ and a contribution
that depends explicitly on the nuclear part
of initial incoming state (\ref{incoming0}). 
The action integral depends only on the electronic levels 
and reads  $\int_{\zeta}\sqrt{2(E-e_2(z)}dz$ where $\zeta$ is a 
loop in the complex plane based at the origin encircling $z_0$. 
The contribution from the nuclear part of the wave packet depends
on the shape of its momentum/energy density. It is that last contribution
that makes the obvious candidate given by the
imaginary part of the action integral taken at 
the classical energy $E$, miss the actual value of the decay rate
$\alpha^*$. 
In that sense, (\ref{molLZ}),
which we could call a molecular Landau-Zener formula,
cannot be determined 
from the usual adiabatic Landau-Zener formula
with just the knowledge of the electronic levels and
the classical nuclear momentum close to the avoided crossing. 
Indeed, our analysis shows that we also need to take 
into account the details of the incoming wave packet
to determine (\ref{molLZ}). 
This is why we resort to coherent
states to get such accurate asymptotics. \\

The way we obtain all our results is by employing a time--independent
scattering theory approach that uses generalized eigenfunctions
of the full Hamiltonian. We expand the wave function in terms of
the generalized eigenfunctions and calculate the large $|t|$ 
asymptotics. For every incoming momentum $k$ there is 
classical energy conservation, but a different
probability of making the non--adiabatic transition. We obtain 
the correct $\alpha^*$ and $\eta_+$ by computing the averages over
$k$ rather than by doing one calculation based on
the average incoming momentum $\eta_-$.

\vspace{.5cm} 
\noindent {\bf Remarks}\\
1.\ We obtain the analogous results
when the incoming
state is associated with the lower electronic level $e_1$, provided
that we keep the average total energy above both the levels.\\
2.\ There are other components of the scattered wave function.
For example, one should expect a reflected wave on the $e_2$ electronic
level and also a reflected wave on the $e_1$ level. We prove that if the
avoided crossing has a sufficiently small gap, then the other components
are exponentially even smaller in $1/\eps^2$ than the 
transmitted non--adiabatic term we compute. 

\vspace{.5cm}
The second situation we describe in this introduction involves the same
set--up as above, but with the Gaussian incoming states replaced by
more general incoming coherent states. This example illustrates the second 
key feature that our analysis demonstrates: even if
the incoming state is not Gaussian, the outgoing non--adiabatic transition
state, generically, is Gaussian to leading order in $\eps$.

For $m=1,\,2,\,\dots$, we define 
\bea\label{ffim}
&&\ffi_m(A,\,B,\,\eps^2,\,a,\,\eta,\,x)\ =\\
&&\qquad\qquad\qquad 2^{-m/2}\,(m!)^{-1/2}\,A^{-m/2}\,(\overline{A})^{m/2}\,
H_m\left(\frac{x-a}{\eps\,|A|}\right)\,
\phi_0(A,\,B,\,\eps^2,\,a,\,\eta,\,x), \nonumber
\eea
where $H_m$ is the $m^{\mbox{\scriptsize th}}$ order Hermite polynomial.

We now replace (\ref{incoming0}) by
\be\label{incomingj}
e^{i(\eta_-^2/2-e_2(-\infty))t/\eps^2}\,
\ffi_m(A_-+iB_-t,\,B_-,\,\eps^2,\,a_-+\eta_-t,\,\eta_-,\,x)\ \phi_2(x).
\ee

Again, up to exponentially small errors, the large positive $t$
asymptotics
of the solution are multiples of $\phi_2(x)$. 
Their leading behavior is determined by the standard 
time--dependent Born--Oppenheimer approximation,
\bea\nonumber
e^{i\theta_1(\eps)}\,e^{i(\eta_1^2/2-e_2(\infty))t/\eps^2}\,
\ffi_m(A_1+iB_1t,\,B_1,\,\eps^2,\,a_1+\eta_1t,\,\eta_1,\,x)\ \phi_2(x),
\eea
where $A_1,\ B_1,\ a_1,\ \eta_1$, and $\theta_1(\eps)$ are the same as in
our
first example.
However, our Theorem \ref{at_last_m}
shows that the leading order asymptotics of the
non--adiabatic component of the wave function for large
positive $t$ again have the form of
a freely propagating Gaussian
\bea\nonumber\hspace{-4mm} 
c_m\ \eps^{-m} 
\ e^{-\alpha^*/\eps^2}\,e^{i\theta_+(\eps)}\,
e^{i(\eta_+^2/2-e_1(\infty))t/\eps^2}\,
\ffi_0(A_{+}+iB_{+}t,\,B_{+},\,\eps^2,\,a_{+}+\eta_{+}t,
\,\eta_{+},\,x)\ \phi_1(x),
\eea
and display a pre-exponential factor of order $\eps^{-m}$.
The values of $\alpha^*$,  $A_{+}$, $B_{+}$, $a_{+}$, and
$\eta_{+}$ are the same as in our first example, and 
we determine the prefactor $c_m$. The numerics 
presented below clearly illustrate these features. 

\vskip 5mm
Our most general result, Theorem \ref{mai}, extends these results in several ways.
First, we
can handle electron Hamiltonians $h(x)$ that are $m\times m$ 
complex hermitian matrices 
which have two levels of interest that have an avoided crossing.
These levels must stay well separated
from the rest of the spectrum of $h(x)$.
Second, we can handle  situations
in which  several levels display certain patterns of avoided
crossings. For example, when two levels have an avoided crossing for one
value 
of $x$,
and one of those
levels has another avoided crossing with a third level
for some other value of $x$.
However, in such cases, we can only study the non--adiabatic
components for certain levels.
The ones we can handle depend on the order in which the levels
have the avoided crossings. 
Third, we can consider more general incoming states that do not
have the form of the $\ffi_j$'s considered above. They are 
characterized by an energy (or momentum) distribution which is 
sharply peaked around some fixed energy, so that a semiclassical 
analysis can be performed. In such general cases also, the nuclear 
part of the non--adiabatic wave function is Gaussian and exponentially 
small, with a decay rate sharing the properties described above. 

\vskip 5mm
The paper is organized as follows:\quad
In the rest of the Introduction, we review the relevant literature and
present numerical results for the above examples.
They show excellent agreement with our analysis. In Section 2, we set up
the general problem we study. We state most of our hypotheses here and make
precise the notion of avoided crossing.
In Section 3, we study generalized eigenvectors of the full
Hamiltonian. In particular, their WKB--type analysis in the complex
plane is performed here.
We superimpose the generalized eigenvectors to generate solutions to the 
time--dependent Schr\"odinger equation and   construct 
asymptotic scattering states in Section 4. 
Non--adiabatic transition asymptotics are studied in Section 5, where our
most general result is stated as Theorem \ref{mai}. 
Further properties and estimates on the energy and momentum shifts are provided
in Section 5.
Section 6 is devoted to the special case of interest where 
the nuclear part of the incoming state is a Gaussian or a Gaussian times 
a Hermite polynomial as in (\ref{incomingj}). Finally, Section 7
contains the proofs of several technical results that are stated in
the earlier sections.

From this outline, one can see that our results depend crucially on the
properties of generalized eigenvectors of the full Hamiltonian. We prove 
these properties by applying the ideas and results of Joye \cite{j}, \cite{joye} 
that provide exponentially accurate WKB--type results in a generic 
avoided--crossing regime, generalizing earlier two--level adiabatic 
techniques from \cite{jkp}, \cite{jp3}, \cite{jp5}. See also 
\cite{mn}, \cite{r} for stationary results of the same kind. 
That a complex WKB-type
analysis plays an important role here should be no surprize. Indeed,
in the ODE context of adiabatic-like problems dealt with in the references above,
the complex WKB approach proved to be the most efficient method providing a
quantitative analysis of the exponentially small leading order term of the
Landau-Zener mechanism. See, however, \cite{hagjoy8} and \cite{bt} for
a different successful approach of such problems, based on optimal
truncation techniques.

There are mathematical results on the exponentially small size 
of non--adiabatic transitions in the Born-Oppenheimer approximation,
and for related problems. See, {\it e.g.},
\cite{hagjoy6}, \cite{marsor}, \cite{benmar}, \cite{mp36},
\cite{mp37}.
However, to the best of our knowledge,
there are no rigorous results on this
topic in the literature that actually compute the leading
asymptotics 
of non--adiabatic transitions in our time--dependent PDE setting. We
have recently learned that Betz and Teufel, \cite{bt0},
are adapting techniques from
\cite{bt} to the Born-Oppenheimer setup.
They have formal and numerical results 
for specific electronic hamiltonians in agreement with ours.
Also, rigorous results on the propagation of wave packets
through avoided crossings, representing first attempts to
unravel the molecular Landau-Zener 
mechanism, are obtained in \cite{hagjoy1}, \cite{hagjoy2}.
(See also \cite{rousse}.)
In those papers, the gap 
$\delta$ shrinks to zero with $\eps$ in such a way that the transitions are 
of order one, so that they can be
computed by perturbation theory.
This is in contrast to 
the present situation, in which $\delta$ is small
but fixed as $\eps\ra 0$,
and the transitions are exponentially small.
  
Because of the importance of the Landau-Zener mechanism to
molecular physics, there are relevant papers in the physics
and chemistry literature. 
See, {\it e.g.}, \cite{coker}, \cite{tully}, \cite{webster}.
\\

\vskip 5mm
\subsection{Numerical Simulations for a Gaussian Initial State}

We now present graphical results of a numerical simulation in which the
initial state is a Gaussian function associated with the upper energy
level for a two level system.
These plots are in very good agreement with the results of our analysis.

We have numerically integrated equation (\ref{schr}) with $\eps=0.2$ for the
Hamiltonian function
$$h(x)\ =\ \frac 12\ \pmatrix{1 &\tanh(x) \cr\tanh(x) & -1}.$$
The energy levels are $\ds\pm\,\frac 12\,\sqrt{1+\tanh(x)^2}$, and 
there is an avoided crossing at $x=0$ with a minimum gap of 1.
The initial state is the eigenvector associated with the upper energy
level times the Gaussian
$\phi_0(A_0+itB_0,B_0,\eps^2,\eta t,\eta,x),$ where $A_0=B_0=\eta=1$,
with the initial time 
$t=-10$. The following two figures show the initial position and
momentum probability densities, respectively.
In both plots, the probablity of being on the lower energy
level is zero.

\vskip 8mm
\centerline{\includegraphics[height=2in,width=6in]{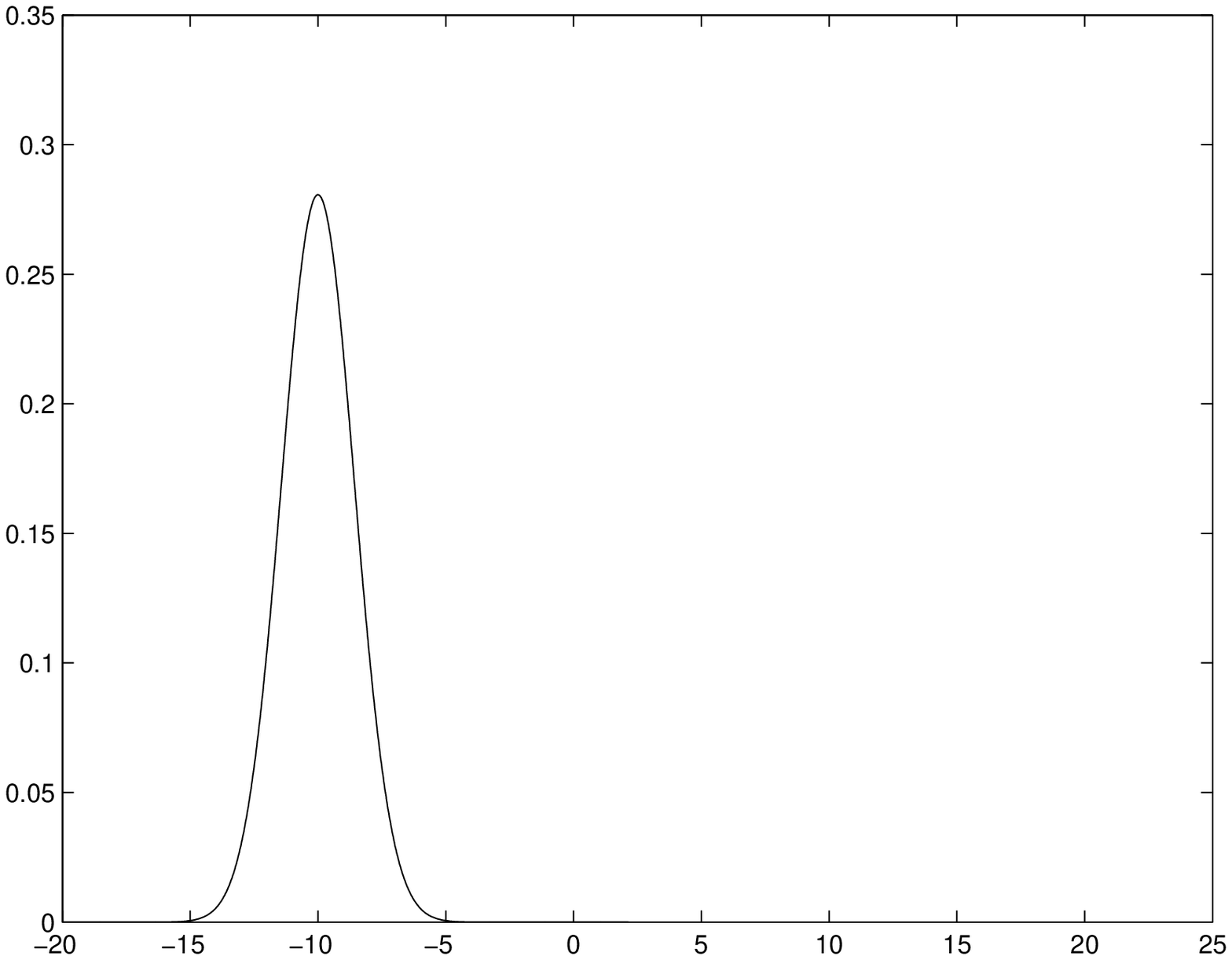}}

\vskip 5mm
\noindent {\bf Figure 2.} Position space plot at time $t=-10$ of the probability
density
for being on the upper energy level (solid line),
and $3\times 10^{8}$ times the
probability density for being on the lower energy level (dotted line).

\vskip 5mm
\centerline{\includegraphics[height=2in,width=6in]{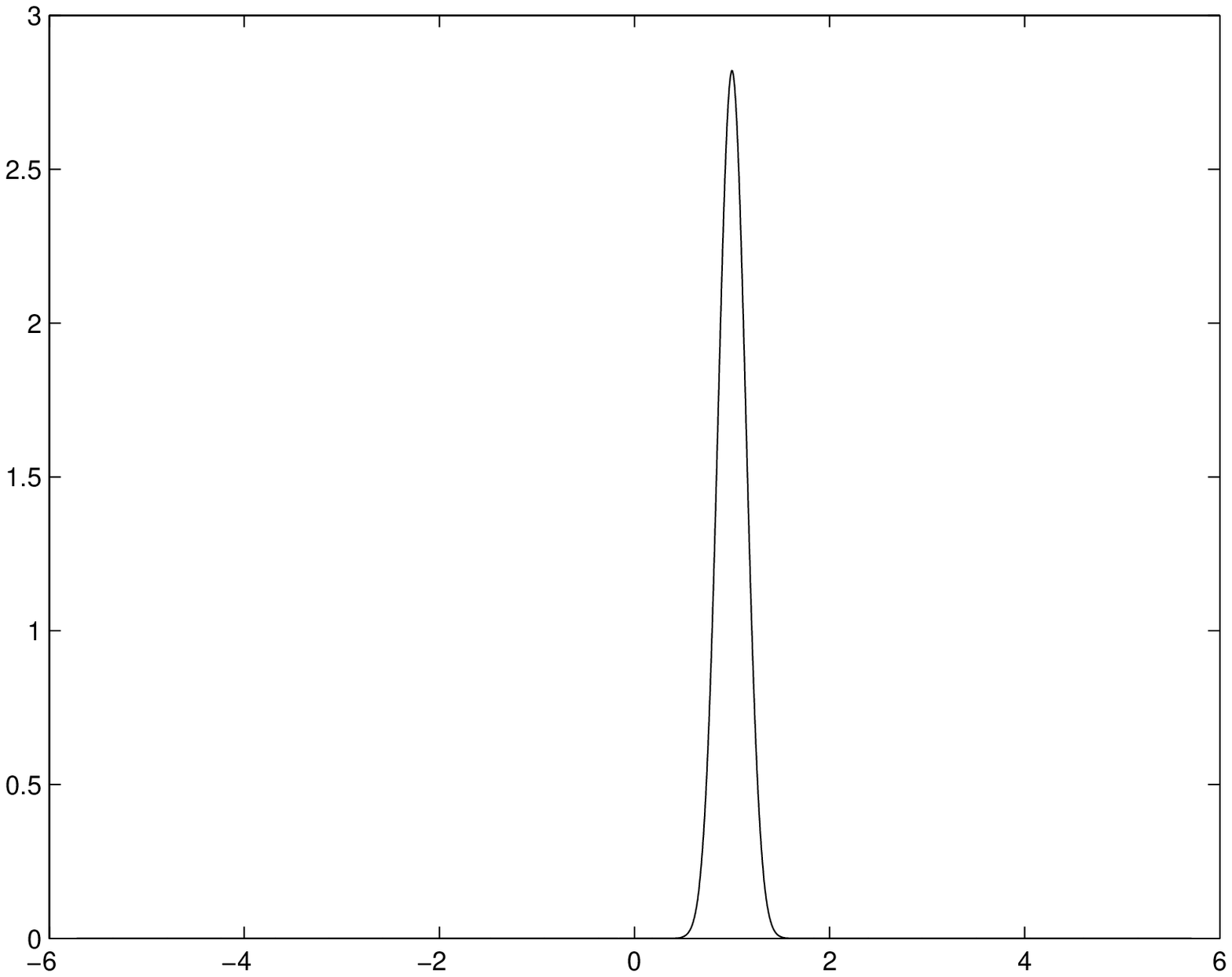}}

\vskip 5mm
\noindent {\bf Figure 3.} Momentum space plot at time $t=-10$ of the probability
density for being on the upper energy level
(solid line), and $3\times 10^{8}$ times the
probabilty density for being on the lower energy level (dotted line).

\vskip 8mm
The following two plots show the position and momentum
probability densities at $t=9$
after the wave function has interacted with the avoided crossing.
The component associated with the lower energy level has mean momentum
2.05. It is evident from the plot that it is greater than 2.

The naive energy conservation calculation predicts the following:
The total energy is $E\ =\ \eta^2/2\,+\,1/2\,\sqrt{1+\tanh(-10)^2}\ =\ 1.2071.$
After the transition to the lower surface, the kinetic energy should be
this value plus $\sqrt{2}/2$, so $\eta_1^2/2\,=\,1.9142$. This predicts a final
momentum after the transition of $\eta_1\,=\,1.9566$.

\vskip 8mm
\centerline{\includegraphics[height=2in,width=6in]{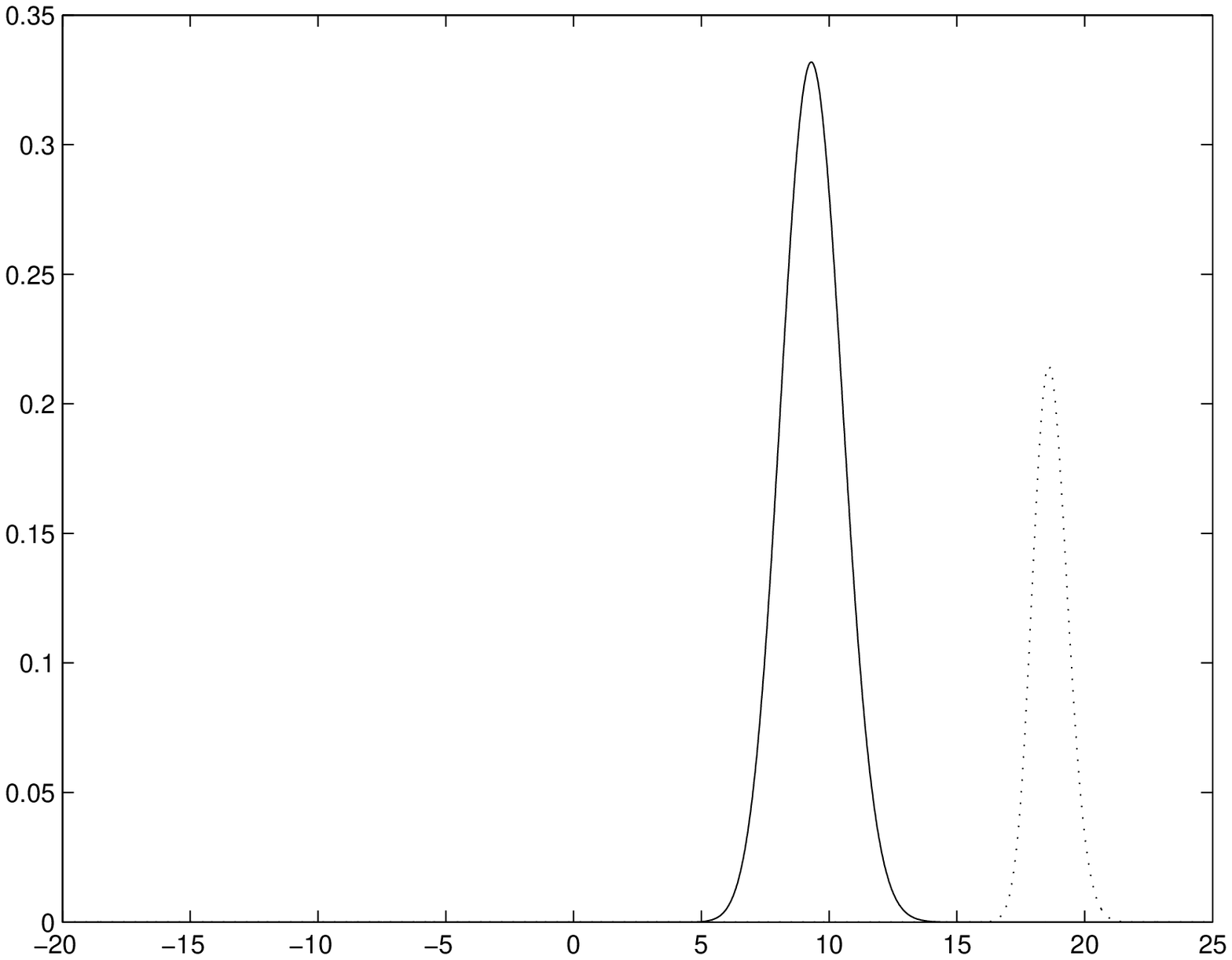}}

\vskip 5mm
\noindent {\bf Figure 4.} Position space plot at time $t=9$ of the probability density
for being on the upper energy level
(solid line), and $3\times 10^{8}$ times the
probabilty density for being on the lower energy level (dotted line).

% gap=1;
%steps=3000;
%t0= -10;
%t1=9;
%dt=(t1-t0)/steps;
%xmax=25;
%xmin= -20;
%xchg=xmax-xmin;
%sizer=2048;
%psi=zeros(sizer,2);
%psihat=psi;
%Phi1=psi;
%Phi2=psi;
%u=zeros(2,2,sizer);
%eta=1;
%plot(x,abs(psiup).^2,'k-',x,3*10^8*abs(psidown).^2,'k:')
%plot(p,3*10^8*abs(fftshift(fft(psidown))).^2*dx/sizer/dp,'k:',
%     p,abs(fftshift(fft(psiup))).^2*dx/sizer/dp,'k-')

\vskip 5mm
\centerline{\includegraphics[height=2in,width=6in]{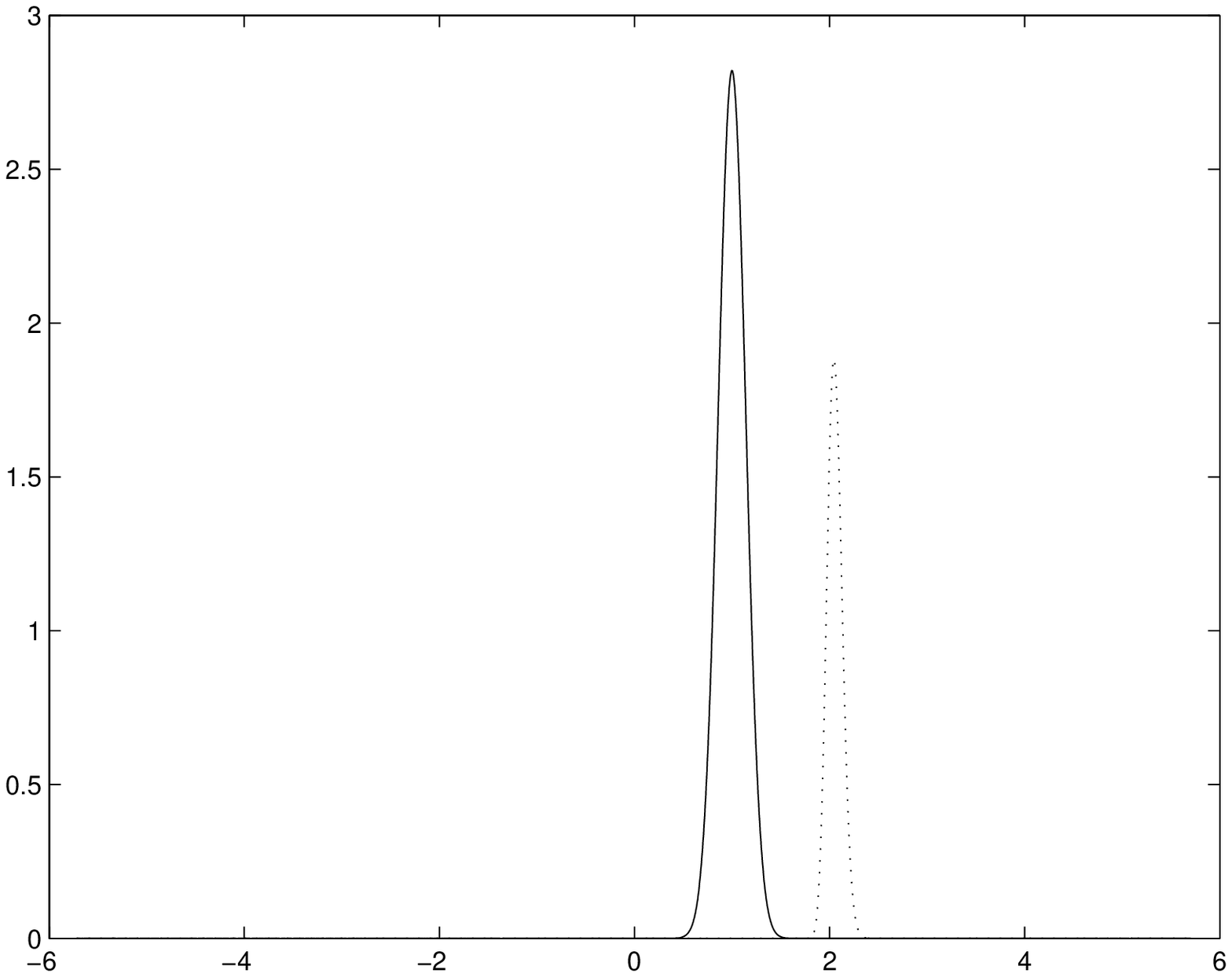}}

\vskip 5mm
\noindent {\bf Figure 5.} Momentum space plot at time $t=9$ of the probability density
for being on the upper energy level (solid line),
and $3\times 10^{8}$ times the 
probabilty density for being on the lower energy level (dotted line).

\vskip 8mm
\subsection{Numerical Simulations for More General Initial States}

We next present the results for the same system as above, but where the initial
Gaussian $\phi_0$ has been replaced by $\phi_3$. See (\ref{ffim}).
Note that the transition
amplitude is significantly larger than in the example above,
and that the component of the wave function
that makes the transition to the lower level is approximately a Gaussian.
The value of epsilon $\eps=0.2$ is not particularly small, so the component of
the final state that does not make a transition is only approximately a $\phi_3$
wave packet. We have chosen this relatively large value of epsilon
to avoid numerical difficulties in integrating equation (\ref{schr}).

We should also note that the naive energy conservation calculation
again predicts that the component of the wave function on the lower
level should have mean momentum 1.9566.
Since initial wave function has a greater momentum
uncertainty than in the Gaussian example above, we see an even
greater discrepancy between this prediction and the correct value.
Our simulation yields a value of roughly 2.25.

\vskip 8mm
\centerline{\includegraphics[height=2in,width=6in]{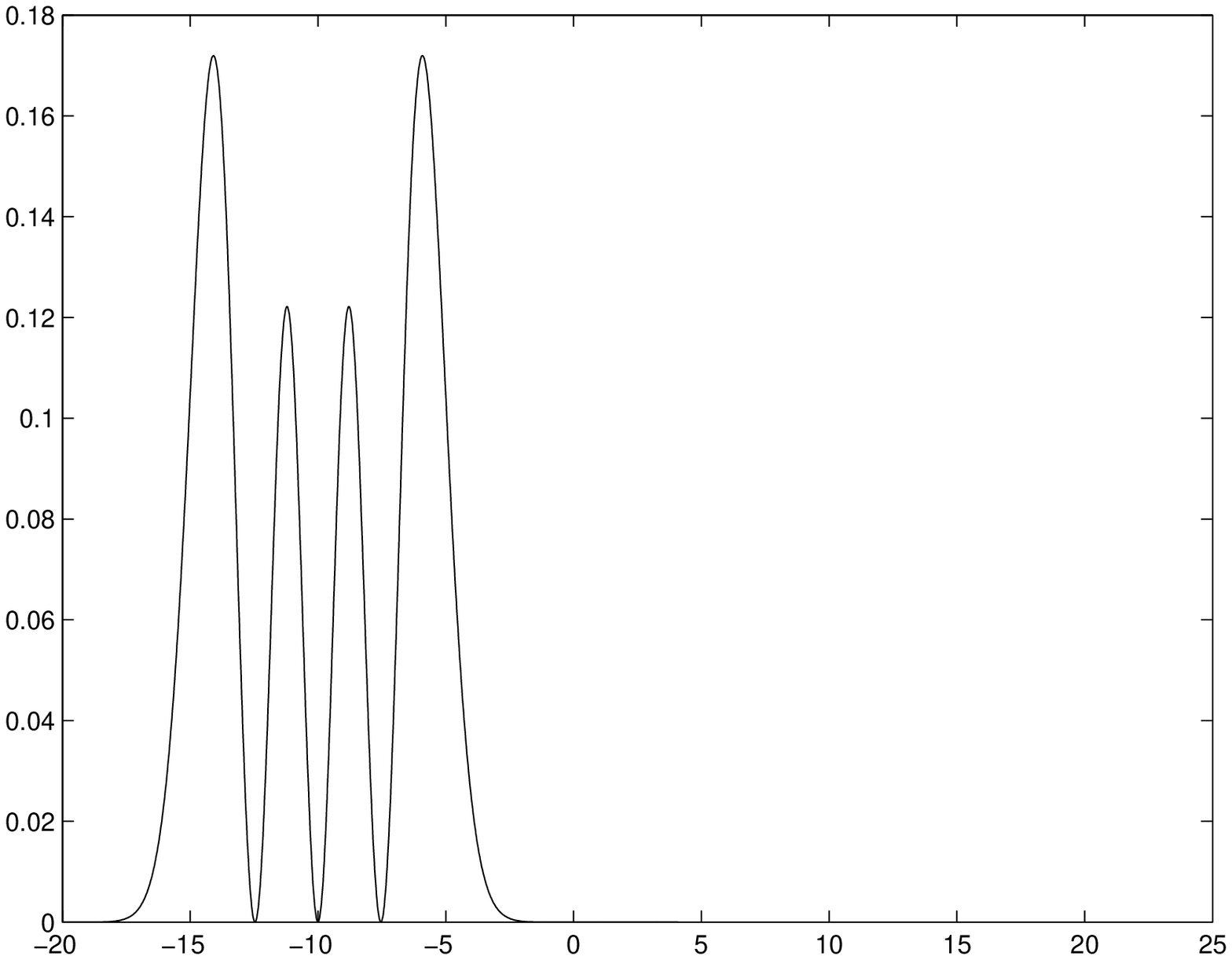}}

\vskip 4mm
\noindent {\bf Figure 6.} Position space plot at time $t=-10$ of the probability
density for being on the upper energy level (solid line),
and $10^{7}$ times the
probabilty density for being on the lower energy level (dotted line).

\vskip 7mm
\centerline{\includegraphics[height=2in,width=6in]{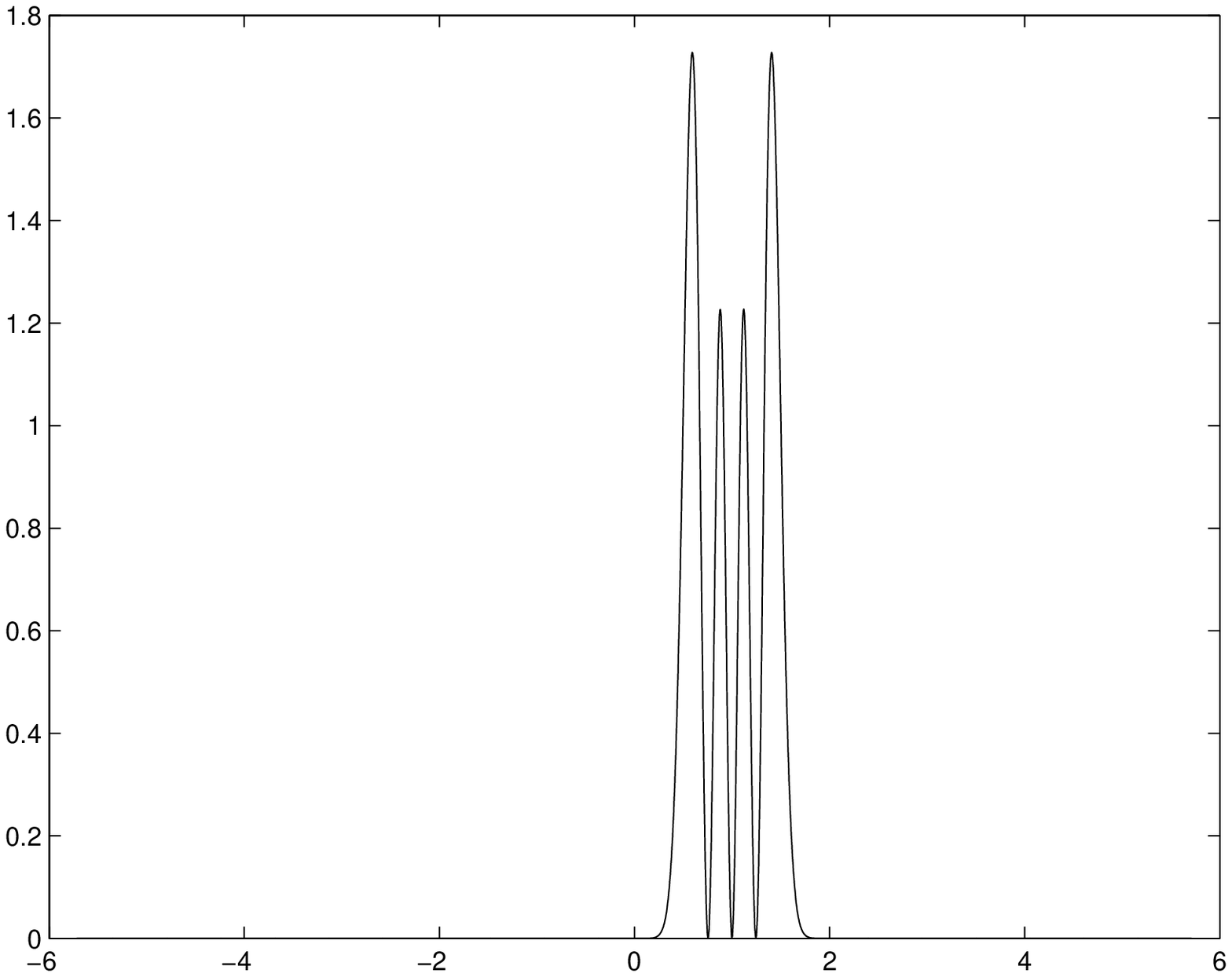}}

\vskip 4mm
\noindent {\bf Figure 7.} Momentum space plot at time $t=-10$ of the probability
densityfor being on the upper energy level (solid line),
and $10^{7}$ times the
probabilty density for being on the lower energy level (dotted line).

\vskip 7mm
\centerline{\includegraphics[height=2in,width=6in]{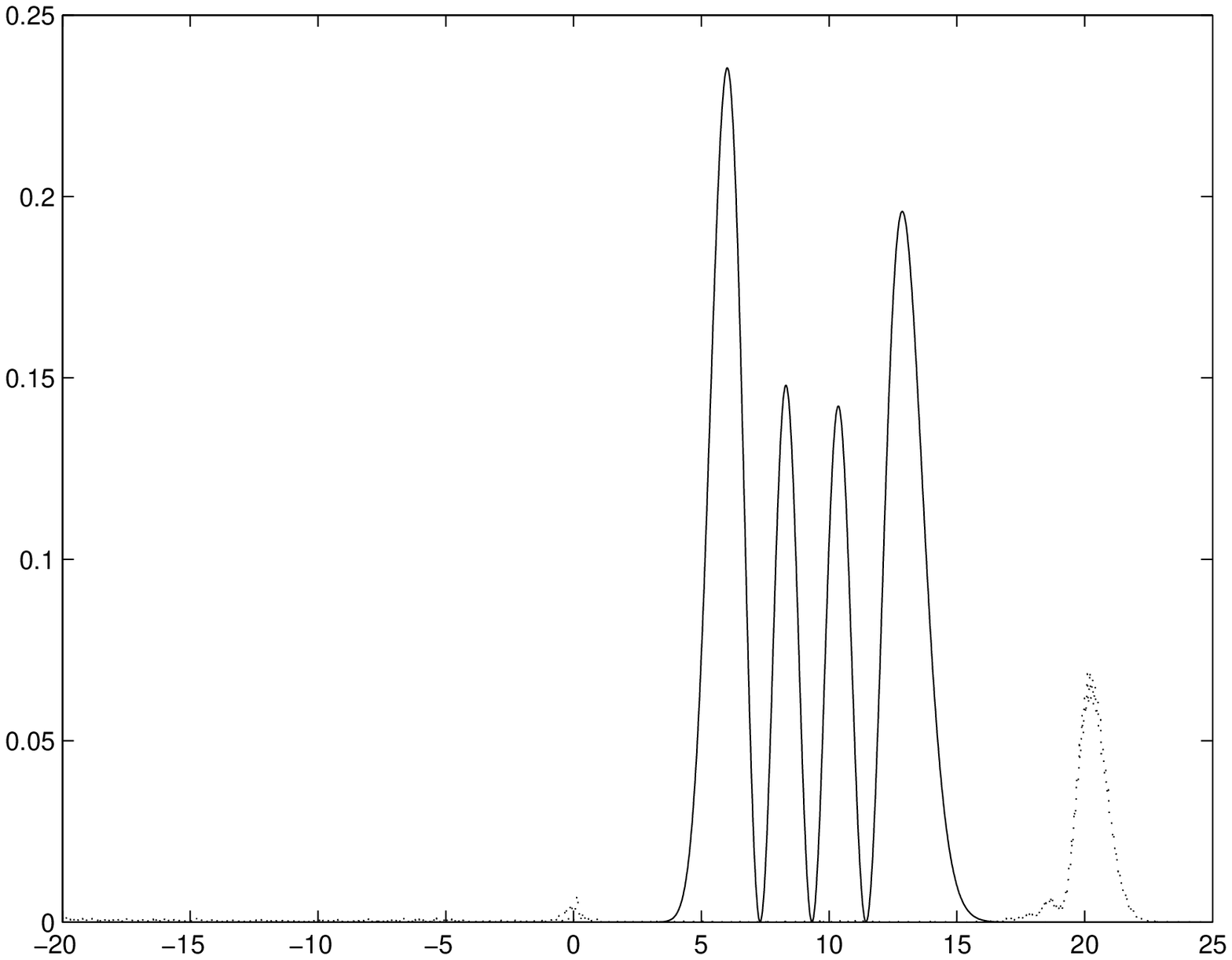}}

\vskip 4mm 
\noindent {\bf Figure 8.} Position space plot at time $t=9$ of the probability density
for being on the upper energy level (solid line),
and $10^{7}$ times the
probabilty density for being on the lower energy level (dotted line).

\vskip 7mm
\centerline{\includegraphics[height=2in,width=6in]{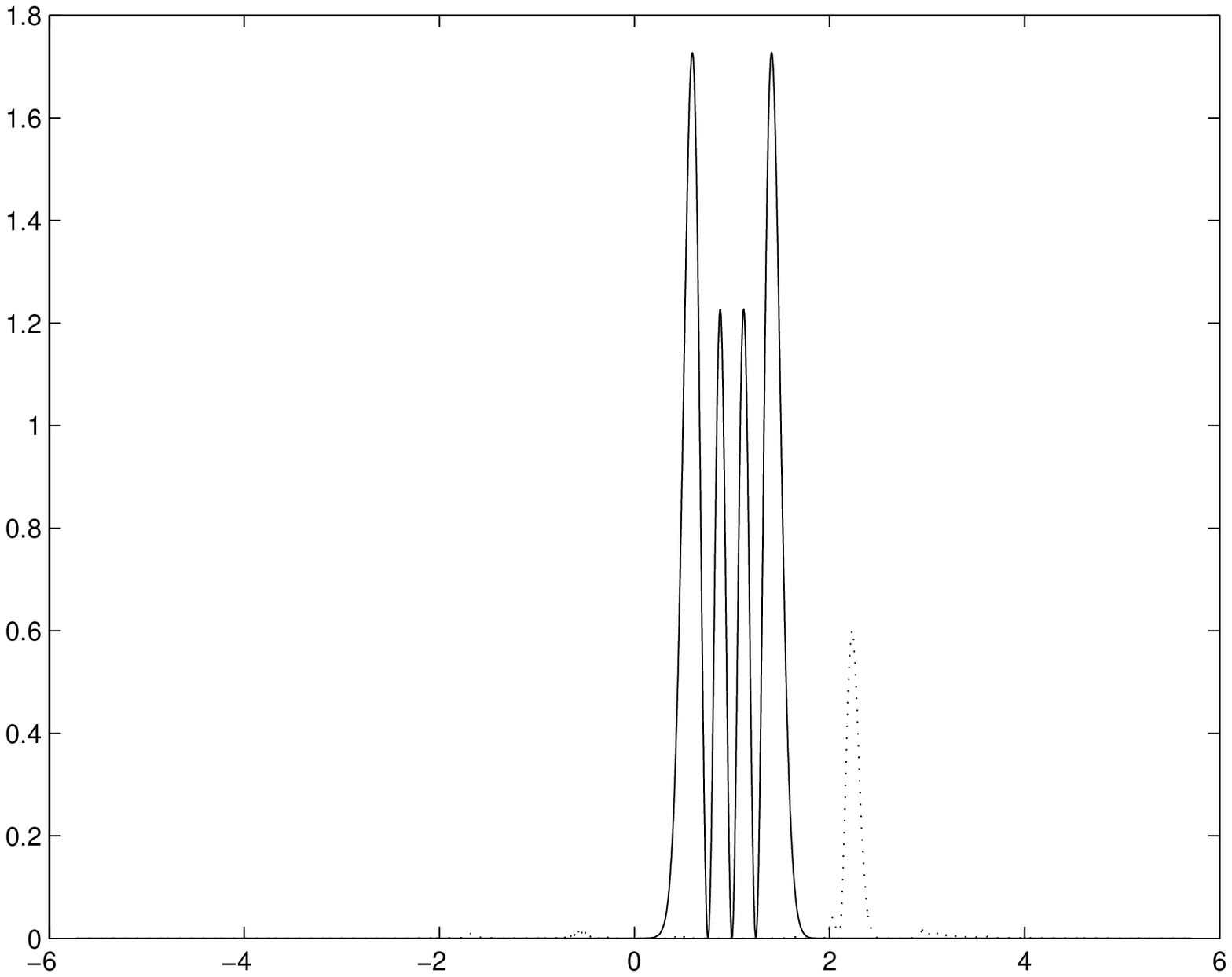}}

\vskip 4mm 
\noindent {\bf Figure 9.} Momentum space plot at time $t=9$ of the probability density
for being on the upper energy level (solid line), and $10^{7}$ times the
probabilty density for being on the lower energy level (dotted line).

\vskip 8mm \noindent
{\bf Acknowledgements}\quad George Hagedorn wishes to thank the Institut Fourier
and the City of Grenoble for their kind hospitality and support during 2003
and 2004 when this research was conducted.

\vskip .3in
\section{Hypotheses for the Electron Hamiltonian}
\setcounter{equation}{0}

We begin with three general assumptions about the electron
Hamiltonian $h$. We then impose two more assumptions that
make precise the avoided crossing situations we can handle.

\vskip 3mm
\noindent
{\bf H1}:\quad We assume $z\mapsto h(z)$ is a $m\times m$ 
matrix--valued analytic function that is analytic in
$z\in \rho_\alpha=\{z=x+iy\,:\,|y|\leq \alpha\}$, 
where $\alpha >0$.
We assume $h(z)$ is self-adjoint for $z\in\R$. 

\vskip 3mm
Since we work in a scattering framework, we further assume:

\vskip 3mm
\noindent
{\bf H2}:\quad There exist $\nu >1/2$, $c$, and two matrices
$h(\pm\infty)$, such that for all $x\in\R$,
$$
\sup_{|y|\leq\alpha}\ \|h(x+iy)-h(\pm\infty)\|\ 
\leq\ \frac{c}{<x>^{2+\nu}},
$$
where $<x>$ denotes $(1+x^2)^{1/2}$.

\vskip 3mm
The rate of convergence
in this assumption can certainly be weakened. 
However, general scattering theory is not the main point
of the present study.

\vskip 3mm
\noindent
{\bf H3}:\quad We assume the spectrum $\sigma(h(x))$ of 
$h(x)$ consists of $m$ non--degenerate eigenvalues 
$$
\sigma(h(x))\ =\ \{e_j(x)\}_{j=1,\cdots, m},
$$
for any $x\in\R \cup\{\pm\infty\}$. 

\vskip 3mm
We let $\phi_j(x)$, $j=1,\cdots,m$, denote the
corresponding eigenvectors, characterized up to constant phases by the
following
conditions
\be\label{elev}
\|\phi_j(x)\|\ \equiv\ 1, \qquad \mbox{and}\qquad
\bra \phi_j(x),\,\phi_j'(x)\ket\ \equiv\ 0, \ \ \ 
\forall\,j=1,\cdots,m,
\ee
where the the prime denotes the derivative with respect to $x$. The eigenvectors
are analytic in some narrow open strip containing the real axis \cite{k}.

By using the Cauchy formula, it is easy to check that
our hypotheses imply
\be\label{elecdecay1}
\frac{d^n}{dx^n}\,\left(\,e_j(x)-e_j(\pm\infty)\,\right)\,\ 
=\ O(<x>^{-(2+\nu)})
\ee
and
\be\label{elecdecay2}\frac{d^n}{dx^n}\,
\left(\,\phi_j(x)-\phi_j(\pm\infty)\,\right)\ =\ O(<x>^{-(2+\nu)}),
\ee
for any $n\in\N$.

\vskip 3mm
We now make specific assumptions concerning
avoided crossings for $h$.

The idea is to assume $h(x)$ belongs to a smooth family of 
electron Hamiltonians $h(x,\delta)$.
When $\delta=0$, we assume there are actual crossings.
When $\delta\neq 0$, we assume there are no crossings for 
real values of $x$.
The electron Hamiltonians we actually use have the form
$h(x,\delta)$ for some small, but fixed value of $\delta$.

Our precise assumption is the following:

\vskip 3mm
\noindent
{\bf H4}:\quad {For each fixed $\delta\in[0,d]$, the matrix
$h(x,\delta)$ satisfies {\bf H1} in a strip $\rho_{\alpha}$
independent of $\delta$, and $h(z,\delta)$ is
$C^2$ as a function of the two variables
$(z,\delta)\in \rho_{\alpha}\times [0,d]$.
Moreover, $h(\cdot)$ satisfies {\bf H2} uniformly for
$\delta\in[0,d]$, with limiting values $h(\pm\infty,\delta)$
that are $C^2$ functions of $\delta\in [0,d]$.}

\vskip 3mm
Again, some of our results hold under weaker smoothness 
assumptions.

\vskip 3mm
We can deal with multiple avoided crossings, but cannot
deal with all possible patterns of avoided crossings.
The following assumption describes the ones we allow:

\vskip 3mm
\noindent
{\bf H5}:\quad For each $x\in\R$ and each $\delta\in[0,d]$,
$\sigma(h(x,\delta))$ consists of $m$ real eigenvalues
\be\label{order}
  \sigma(h(x,\delta))\ =\ \{e_1(x,\delta),\,
e_2(x,\delta),\,\cdots,\,e_m(x,\delta)\}\ \subset\ \R.
\ee
When $\delta>0$ we assume these are distinct 
for $x\in [-\infty,\,+\infty]$ and are labeled by
$$
e_1(x,\delta)< e_2(x,\delta)<\cdots <e_m(x,\delta).
$$
When $\delta =0$, the eigenvalues are $m$ analytic functions that
have finitely many real crossings at
$x_1\leq x_2\leq \cdots \leq x_p$, with $p\geq 1$.
We assume the eigenvalues have $m$ distinct limits as $x\to -\infty$
and as $x\to\infty$.
We label these eigenvalues $e_j(x,0)$ in a way that is discontinuous
in $\delta$ near $\delta=0$. This labeling is determined by the following
conditions:\\
i) For all $x<x_1$,
$$
  e_1(x,0)< e_2(x,0)<\cdots <e_m(x,0).
$$
ii) For all $j<l\in\{1,2,\cdots ,n\}$,
there exists at most one $x_r$ with
$$
  e_j(x_r,0)\,-\,e_l(x_r,0)\ =\ 0,
$$
and if such an $x_r$ exists, we have
\be\label{gc}
\frac{\partial\phantom{i}}{\partial x}
  \left(e_j(x_r,0)-e_l(x_r,0)\right)\ >\ 0.
\ee
iii) For all $j\in\{1,2,\cdots ,n\}$,
the eigenvalue $e_j(x,0)$ crosses eigenvalues whose
indices are all superior to $j$ or all inferior to $j$.

\vskip 3mm
\noindent{\bf Remarks:}\\
i) The parameter $\delta$ can be understood as a
coupling constant that controls the strength of the
perturbation that lifts the degeneracies of $h(x,0)$
on the real axis.\\
ii) Analyticity of the eigenvalues $e_j(x,0)$
on the real axis follows from the self--adjointness
of $h(x,0)$.\\
iii) The crossings are assumed to be generic
in the sense that the derivatives of $e_j-e_k$ are non-zero at the crossing
$x_r$.
This ensures that when $\delta>0$ is small, the generic behavior
(\ref{generic})
holds at the corresponding complex crossing points.\\
iv) When $m=2$, {\bf H5} requires that the two eigenvalues have exactly
one generic crossings when $\delta=0$.\\
v) The crossing points $\{x_1,x_2,\cdots ,x_p\}$ need not be distinct,
which is important when
the Hamiltonian possesses symmetries. However, for each $j=1,\cdots, n$, the
eigenvalue
$e_j(x,\delta)$ experiences avoided crossings with $e_{j+1}(x,\delta)$
and/or
$e_{j-1}(x,\delta)$ at a subset of distinct points
$\{x_{r_1},\cdots ,x_{r_j}\}\subseteq \{x_1,x_2,\cdots ,x_p\}$.

\vskip 3mm
For certain results, we also impose the condition that
these avoided crossings be generic
in the sense of \cite{oldHag} and \cite{j}.
This condition essentially says that the low order Taylor series
coefficients of certain quantities do not vanish at the
crossing when $\delta=0$. 

\vskip 3mm
\noindent
{\bf H6}:\quad
Near an avoided crossing
of $e_j(x,\delta)$ and $e_n(x,\delta)$, 
there exist  $a>0$, $b>0$, and $c\in\R$, such that 
\be\label{genericstuff}
e_n(x,\delta)\,-\,e_j(x,\delta)\ =\
\pm\ \sqrt{ax^2\,+\,2cx\delta\,+\,b^2\delta^2\,+\,R_3(x,\delta)},
\ee
where $c^2<a^2b^2$ and $R_3(x,\delta)$ is a remainder of order $3$ in
$(x,\delta)$ close to $(0,0)$.

\vskip 3mm
Our final hypothesis involves both the electron Hamiltonian and
an interval of energies, $\Delta$. We ultimately consider states
of the full Hamiltonian whose energy is concentrated in 
$\Delta$, with $\Delta$ high enough that scattering
onto all the electron energy levels is possible.
An energy range that satisfies this condition can always be chosen for
some strip $\rho_\alpha$, provided the minimum value in $\Delta$ is large
enough.

\vskip 3mm
\noindent
{\bf H7}:\quad The interval $\Delta\in\R$ is compact and has non-empty
interior.
Furthermore, it is chosen so that
$$
\inf_{E\in\Delta \atop {z\in\rho_\alpha \atop \delta \in [0,\delta]}}
\,|E-e_j(z,\delta)|\ >\ 0.
$$

\vskip 3mm
\section{Generalized Eigenvectors}
\setcounter{equation}{0}

For energies $E\in\Delta$, we
construct generalized eigenvectors for the full Hamiltonian. For the time
being, the parameter $\delta>0$ is fixed and we drop it in the notation. 
The generalized eigenvectors are solutions $\Psi(x,E,\eps)\in\C^m$
to the time--independent Schr\"odinger equation
\be\label{schsta}
\left(\,-\,\frac{\eps^4}{2}\,\frac{\partial^2}{\partial x^2}\,+\,h(x)\,\right)
\ \Psi(x,E,\eps)\ =\ E\ \Psi(x,E,\eps).
\ee
For each
$E\in\Delta$, the set of such solutions is $2m$-dimensional, and 
individual solutions can be characterized
by their asymptotics at $x=-\infty$ (or at $x=\infty$).

Let
$$
\Phi(x,E,\eps)\ =\
\pmatrix{\Psi(x,E,\eps)\vspace{1.5mm} \cr\vspace{1mm}
i\,\eps^2\,\frac{\partial}{\partial x}\,\Psi(x,E,\eps)}
\ \in\  \C^{2m}.
$$
Then (\ref{schsta}) is equivalent to
\be\label{adiab}
i\,\eps^2\,\frac{\partial}{\partial x}\Phi(x,E,\eps)\
=\ H(x,E)\,\Phi(x,E,\eps),
\ee
where
\bea\label{adiabH}
&& H(x,E)\ =\ \pmatrix{ \0 & \un\ \vspace{1.5mm}\cr 2\,(E\ \un\,-\,h(x))&
\0\ }\
\in\ M_{2m}(\C)
\qquad\quad \mbox{and}  \nonumber\\[3mm]
&& E \ \un\,-\,h(x)\ >\ 0,\quad \mbox{for all}\quad
x\in\R\cup\{\pm\infty\}.
\eea
Here, $\un$ denotes the identity matrix in $\C^m$.
Note that the matrix $H(x,E)$ is not self--adjoint, but satisfies the
relation
\be\label{sympl}
H(x,E)\ =\ J\ H^*(x,E)\ J, \qquad \mbox{where}\qquad
J=\pmatrix{\0 & \un \cr \un & \0}.
\ee

The small $\eps$ asymptotics of solutions to (\ref{adiab}) are studied in
\cite{joye}. Of particular importance to us is Section 7 of \cite{joye}, 
which is devoted to the computation
of exponentially small elements of the related $S$-matrix that we describe
below.
We apply the results of \cite{joye} to (\ref{adiab}),
keeping track of the dependence on $E$.

By our hypotheses on $E$ and $h(x)$, the spectrum of $H(E,x)$ consists
of $2m$ distinct real eigenvalues
\bea\label{sigH}
&&\sigma(H(x,E))\ =\ \{\,k_j^\tau(x,E)\,\}^{\tau=+,-}_{j=1,\cdots, m},
\qquad \mbox{with} \nonumber \\[3mm]
&&k_j^\tau(x,E)\ =\ \tau\,k_j(x,E)\ =\ \tau\,\sqrt{2\,(E-e_j(x))}\
\in\,\R.
\eea
Note that the $k_j^\tau$'s correspond to the classical
momenta associated with the classical potentials $e_j(x)$.

A set of corresponding eigenvectors $\{\,\chi_j^\tau(x,E)\,\}$
is given (in block notation) by
\be\label{chi}
\chi_j^\tau(x,E)\ =\
\pmatrix{\phi_j(x)\vspace{1mm}\cr k_j^\tau(x,E)\,\phi_j(x) }\ \in\
\C^{2m}.
\ee
From these we produce new eigenvectors 
\be\label{ffi}
\ffi_j^\tau(x,E)\ =\ \chi_j^\tau(x,E)\ \frac{1}{\sqrt{2\,k_j(x,E)}}
\ee
that satisfy the normalization convention (\ref{normal}) below,
that wasadopted in \cite{joye}.
This normalization is motivated by the following:
We can write
$$
H(x,E)\ =\sum_{j,\tau}\ k_j^\tau(x,E)\ P_j^\tau(x,E),
$$
where $\{P_j^\tau(x,E)\}$ denotes a set of non-orthogonal projections onto
the eigenspaces of $H(x,E)$. If we define
\be\label{thetav}
\theta_j^\tau(x,E)\ =\ \frac{1}{2}\
\pmatrix{\phi_j(x)\vspace{1mm}\cr \frac{\tau}{k_j(x,E)}\,\phi_j(x) },
\ee
then it is easy to check that
\be\label{proj}
P_j^\tau(x,E)\ =\ |\,\chi_j^\tau(x,E)\,\ket\,\bra\,\theta_j^\tau(x,E)\,|,
\ee
where we have used the bra-ket notation relative to the scalar product in
$\C^{2m}$.
We use the same notation for scalar products in $\C^m$ and $\C^{2m}$,
since no
confusion should arise.

We now see that the eigenvectors (\ref{ffi}) satisfy the normalization
conditions
\be\label{normal}
\left\{\
\matrix{P_j^\tau(x,E)\ \frac{\partial}{\partial
x}\,\ffi_j^\tau(x,E)\,\equiv\,0,
\qquad\qquad\mbox{and}\vspace{3mm}\cr
\bra\,\ffi_j^\tau(0,E),\  J\,\ffi_j^\tau(0,E)\,\ket\ \equiv\ \tau\ \in\ \{
-1,1\}. }\right.
\ee

\vskip .5cm
We note that $H,\ k_j^\tau,\ \chi^\tau_j,\ P^\tau_j,\ \mbox{and}\
\ffi_j^\tau$
are analytic functions of $x$ and $E$ when these variables are in
a neighborhood of $\R\times\Delta$.
More precisely, if $\Delta=[E_1,E_2]$, we define
$D_\beta=\{z\in\C\,:\ \mbox{dist}(z,\,\Delta)<\beta \}$, and these functions are
analytic in $\rho_\alpha\times D_\beta$, for $\alpha$ and $\beta$ small
enough. Here $\alpha$
must be chosen small enough
so that $e_j$ and $\phi_j$ are analytic in $\rho_\alpha$,
(see \cite{k}), and $\beta$ must be small enough so that $|E-e_j(x)|>0$ in
$\rho_\alpha\times D_\beta$.

We later make use of larger values of $\alpha$ in order to take
advantage of the generic multivaluedness of $e_j$ and $\phi_j$ as
functions of $x$.

\vskip .5cm
From \cite{joye}, we now see that any solution to (\ref{adiab})
can be written as
\be\label{solad}
\Phi(x,E,\eps)\ =\ \sum_{j, \tau}\ c_j^\tau(x,E,\eps)\
e^{-\,i\,\int_0^x\,k_j^\tau(y,E)\,dy/\eps^2}\
\ffi_j^\tau(x,E),
\ee
where the scalar coefficients $c_j^\tau\in \C$ satisfy the equation
\be\label{coefD}
\frac{\partial}{\partial x}\,c_j^\tau(x,E,\eps)
\ =\ \sum_{l,\sigma}\ a^{\tau \sigma}_{j l}(x,E)\
e^{i\,\int_0^x\,(\tau k_j(y,E)-\sigma k_l(y,E))\,dy/\eps^2}\
c_l^\sigma(x,E,\eps),
\ee
with
$$
 a^{\tau \sigma}_{j l}(x,E)\ =\
 -\ \frac{\bra\,\ffi_j^\tau(x,E),\ P_j^\tau(x,E)\,
\frac{\partial\phantom{i}}{\partial x}\,
\ffi_l^\sigma(x,E)\,\ket}{\|\,\ffi_j^\tau(x,E)\,\|^2}.
$$
We can rewrite (\ref{coefD}) as an integral equation
\bea\nonumber\hspace{-1cm}
c_j^\tau(x,E,\eps)&=&c_j^\tau(x_0,E,\eps)\\[3mm]
&&+\quad 
\int_{x_0}^x\ \sum_{l,\sigma}\ a^{\tau \sigma}_{j l}(x',E)\
e^{i\,\int_0^{x'}\,(\tau k_j(y,E)-\sigma k_l(y,E))\,dy/\eps^2}\
c_l^\sigma(x',E,\eps)\ dx'.\label{coef}
\eea
As we shall soon see, our hypotheses imply the existence of the
limits
$\lim_{x\pm\infty}\ c_j^\tau(x,E,\eps)\,=\,c_j^\tau(\pm\infty,E,\eps)$, so
that with the notation
$$
{\bf c}^\tau(x,E,\eps)\ =\
\pmatrix{c_1^\tau(x,E,\eps)\vspace{1mm}\cr
c_2^\tau(x,E,\eps)\vspace{1mm}\cr
      \vdots \cr c_m^\tau(x,E,\eps)}\ \in\ \C^m,
$$
we can define an associated $S$--matrix, ${\cal S}\in M_{2m}(\C)$, by the
identity
\be\label{scoef}
  {\cal S}(E,\eps)\ 
\pmatrix{ {\bf c}^+(-\infty,E,\eps)\vspace{1mm}\cr {\bf
c}^-(-\infty,E,\eps)}\ =\
\pmatrix{ {\bf c}^+(+\infty,E,\eps)\vspace{1mm}\cr {\bf
c}^-(+\infty,E,\eps)}.
\ee
This $S$--matrix naturally takes the block form
\be\label{bs}
{\cal S}(E,\eps)\ =\
\pmatrix{{\cal S}^{++}(E,\eps) &{\cal S}^{+-}(E,\eps)\vspace{2mm}\cr
{\cal S^{-+}}(E,\eps) &{\cal S}^{--}(E,\eps)}.
\ee
Due to the symmetry (\ref{sympl}), it also satisfies the
relation (see \cite{joye}),
$$
{\cal S}^{-1}(E,\eps)\ =\ R\ {\cal S}^*(E,\eps)\ R,\qquad
\mbox{where}\qquad
R\ =\ \pmatrix{\un & \phantom{-}\0 \cr \0 &-\un}.
$$
Its elements describe transmission and reflection coefficients at fixed
energy $E$ which play key roles in our analysis. The
off-diagonal elements are exponentially small and their asymptotics are
determined in \cite{joye}.

With this notation, the generalized eigenvectors are given by
\bea\label{wkb0}
\Psi(x,E,\eps)&=&\sum_j\ \frac{1}{\sqrt{2\,k_j(x,E)}}\ \phi_j(x)\\[3mm]
& &\qquad\times\quad \left(\,c_j^+(x,E,\eps)\,
e^{-\,i\,\int_0^x\,k_j(y,E)\,dy/\eps^2}\ +\
c_j^-(x,E,\eps)\,e^{i\,\int_0^x\,k_j(y,E)\,dy/\eps^2}\,\right).\nonumber
\eea
Since
$\int_0^x\,k_j(y,E)\,dy\ \simeq\ 
x\,k_j(\pm \infty,E)\ =\ x\,\sqrt{2\,(E-e_j(\pm\infty))}$
as $x\ra \pm \infty$,
the component of (\ref{wkb0}) that describes
a wave traveling from the left to the right is
labeled by $-$, and the component that describes a wave traveling from
the right to the left is labeled by $+$. Note also that (\ref{wkb0}) is
simply a WKB decomposition of the generalized eigenvectors.

\vskip .5cm
We now state some of the general properties of the coefficients
$c_j^\tau(x,E,\eps)$ and of the phases
$e^{i\,\int_0^x\,k_j(y,E)\,dy/\eps^2}$ that allow us to justify
the scattering results described above.

\begin{lem}\label{asesx}
Our hypotheses on $h(x)$ imply the following,
uniformly for $E\in\Delta$ and all $n\in\N$:
\bea\label{consder}
&&0\ <\ C_1(n)\ \leq\ \frac{\partial^n}{\partial E^n}\,k_j(x,E)\
\leq\ C_2(n)\ <\ \infty,\qquad\mbox{and}\\
&&\frac{\partial^n}{\partial E^n}\,(k_j^\sigma(x,E)-k_j^\sigma(\pm\infty,
E))
\ =\ O(<x>^{-(2+\nu)}),\qquad \mbox{as}\quad x\ra\pm\infty.
\eea
Thus, if we define $\omega^\sigma_j(\pm\infty,E)\ =\
\int_0^{\pm\infty}\,(k_j^\sigma(y,E)-k_j^\sigma(\pm\infty,E))\,dy$,
we further have
$$
\int_0^x\,k_j^\sigma(y,E)\,dy\ =\ x\,k_j^\sigma(\pm\infty,E)\
+\ \omega^\sigma_j(\pm\infty,E)\ +\ r_j^\sigma(\pm, x,E)
$$
where, uniformly in $E$ and for all $n\in\N$,
$$
\frac{\partial^n}{\partial E^n}\,r_j^\sigma(\pm,x,E)
\ =\ O(<x>^{-(1+\nu)}),\qquad
\mbox{as}\qquad x\ra\pm\infty.
$$
Moreover, the limits $c_j^\sigma(\pm\infty,E,\eps)$ as $x\ra\pm\infty$
exist, and as $|x|\ra\infty$,
$$
\frac{\partial^n}{\partial E^n}\,c_j^\sigma(x,E,\eps)\ =\
O(1),\quad\mbox{for}\quad
n=0,\,1,
$$
uniformly for $E\in\Delta$.
Also, as $x\ra\pm\infty$ and uniformly for $E\in\Delta$,
\bea\nonumber
&&c_j^\sigma(x,E,\eps)\ -\ c_j^\sigma(\pm\infty,E,\eps)\ =\
O(<x>^{-(1+\nu)}),\qquad\mbox{and}
\\[2mm]\nonumber
&&\frac{\partial}{\partial E}\,
\left(c_j^\sigma(x,E,\eps)\,-\,c_j^\sigma(\pm\infty,E,\eps)\right)
\ =\ O(<x>^{-\nu}).
%\\[2mm]
%&&\frac{\partial^2}{\partial E^2}\,
%\left(c_j^\sigma(x,E,\eps)\,-\,c_j^\sigma(\pm\infty,E,\eps)\right)
%\ =\ O(<x>^{-\nu}).
\eea
\end{lem}

\noindent
{\bf Remarks:}\\
{\bf 1.}\quad This lemma is proved in Section \ref{technical}.\\
{\bf 2.}\quad
Some error terms in the lemma may depend on $\eps$ in a singular
way as $\eps\ra 0$. This will not matter because they will
vanish in the scattering framework we adopt.\\

\subsection{Complex WKB Analysis}
All the information about
transmissions and transitions among the asymptotic eigenstates of the
electronic Hamiltonian is contained in the asymptotic values
of the coefficients $c_j^\sigma(x,E,\pm\infty)$ defined by (\ref{coef}),
and hence, in the matrix ${\cal S}(E,\eps)$.
We extract this information by mimicking the complex WKB method of
\cite{joye}, while keeping track of the $E$ dependence.

The complex WKB method requires hypotheses on the behavior in the
complex plane of the so-called Stokes lines for the equation (\ref{adiab})
in order to provide the required asymptotics.
These hypotheses are global in nature, and in general, are extremely
difficult to check. However, in the physically interesting situation 
of ``avoided crossings,'' they can be easily checked.
We restrict our attention to these avoided crossing situations that are
described below.

\vskip .5cm
We consider the coefficients $c_j$ that are uniquely defined by the
conditions
$$
c_j^\tau(-\infty,E,\eps)\ =\ 1,\quad\mbox{and}\quad
c_k^\sigma(-\infty,E,\eps)\ =\ 0,\quad\mbox{for all}\quad
(k,\,\sigma)\ \neq\ (j,\,\tau).
$$

The key to the complex WKB method lies in the multivaluedness of the
eigenvalues
and eigenvectors of the analytic generator $H(x,E)$ of (\ref{adiab}) in
the complex
$x$ plane. 
For any fixed $E\in\Delta$, $H(\cdot ,E)$ is analytic in $\rho_\alpha$,
and the solutions
(\ref{solad}) to (\ref{adiab}) are analytic in $x$ as well.
However, the eigenvalues and eigenvectors may have branch points in
$\rho_\alpha$ whose
properties are inherited from those of  
the eigenvalues and eigenvectors of $h(\cdot)$. 

Analytic perturbation theory as described in \cite{k} states that the
eigenvalues and eigenprojections of $h(x)$ for real $x$
are analytic on the real axis and admit analytic multivalued
extensions to  $\rho_\alpha$.  The analytic continuations of the
eigenvalues have branch points that are located on
a set of crossing points
$$
\Omega=\{z_0\in\rho_\alpha\setminus \R \,:\,e_j(z_0)=e_k(z_0)\ \mbox{for
some}
\ j,\,k\
\mbox{and some analytic continuations}\}.
$$
Recall that for $\delta=0$, the eigenvalues are analytic at
any crossing points on the real axis.
This follows from the self--adjointness of $h(\cdot)$ on the real axis.
Note also that $\overline{\Omega}=\Omega$ by the Schwarz reflection
principle.

Generically, at a complex crossing point $z_0\in\Omega$, 
we have the following local behavior, where $c\in\C$ is some constant
\be\label{generic}
e_j(z)\,-\,e_k(z)\ \simeq\ c\,(z-z_0)^{1/2}\ (1+O(z-z_0)).
\ee
The eigenprojectors of $h(x)$ also admit multivalued extensions in
$\rho_\alpha\setminus\Omega$, but they diverge at generic eigenvalue
crossing points.
We only have to deal with generic crossing points.

To see what happens to a multivalued function $f$ in
$\rho_\alpha\setminus\Omega$
when we turn around a crossing point, we adopt the following convention:
We denote by
$f(z)$ the analytic continuation of $f$ defined in a neighborhood of the
origin
along some path from $0$ to $z$. Then we perform the analytic continuation
of $f(z)$
along a negatively oriented loop that surrounds only one point
$z_0\in\Omega$.
We denote by $\tilde{f}(z)$ the function we get by coming back to the
original
point $z$.

We define $\zeta_0$ to be a
negatively oriented loop, based at the origin,
that encircles only $z_0$ when $\Im z_0 >0$. When $\Im z_0 <0$, we choose
$\zeta_0$ to be positively oriented.

We now fix $z_0$ with $\Im z_0>0$.
If we analytically continue
the set of eigenvalues $\{e_j(z)\}_{j=1}^m$, along a negatively
oriented loop around $z_0\in\Omega$, we get the set
$\{\widetilde{e}_j(z)\}_{j=1}^m$ with
\bea\nonumber
  \widetilde{e}_j(z)=e_{\pi_0(j)}(z),\quad\mbox{for}\quad j=1,\cdots,m,
\eea
where
\be\label{pi0}
  \pi_0 :\;\{1,2,\cdots,m\}\ra \{1,2,\cdots,m\}
\ee
is a permutation that depends on $z_0$.
As a consequence, the eigenvectors (\ref{elev})
possess multivalued analytic extensions
in $\rho_{\alpha}\backslash \Omega$. 
The analytic
continuation $\widetilde{\phi}_j(z)$ of $\phi_j(z)$
along a negatively oriented loop around
$z_0\in\Omega$, must be proportional to
${\phi}_{\pi_0(j)}(z)$.
Thus, for $j=1,2,\cdots,m$, there exists $\theta_j(\zeta_0)\in\C$, such
that
\be\label{theta}
\widetilde{\phi}_j(z)\ =\ e^{-i\theta_j(\zeta_0)}{\phi}_{\pi_0(j)}(z).
\ee

\vskip 4mm
We now turn from $h(x)$ to $H(x,E)$.

From Hypothesis {\bf H7}, (\ref{sigH}), and (\ref{ffi}), we see that
the set of crossing points for the eigenvalues $\pm k_j(x,E)$ of
$H(x,E)$ is independent of $E$ and coincides with $\Omega$.

Moreover, for $j=1,\cdots,m$, we have
$$
\widetilde{k}_j^\tau(z,E)\ =\ k^\tau_{\pi_0(j)}(z,E),\qquad
\widetilde{\ffi}_j^\tau(z,E)\ =\
e^{-i\theta_j(\zeta_0)}\,{\ffi}_{\pi_0(j)}^\tau(z,E),
$$
where the prefactor $e^{-i\theta_j(\zeta_0)}$ is independent of $E$.

The above implies a key identity for the analytic extensions of
the coefficients $c_j^\tau(z,E,\eps)$, $z\in \rho_{\alpha}\backslash \Omega$.
Since the solutions to (\ref{adiab}) are analytic for all
$z\in\rho_\alpha$, the coefficients $c_j^\tau$ must also be multi-valued.
In our setting, Lemma 3.1 of \cite{joye} implies the following lemma.

\begin{lem}\label{ECHCO}
For any $j=1,\cdots,m$ and $\tau=+,-$, we have
\be\label{echco}
\widetilde{c}^{\tau}_j(z,E,\eps)\
e^{-\,i\,\int_{\zeta_0}\,k^\tau_j(u,E)\,du/\eps^2}\
e^{-\,i\,\theta_j(\zeta_0)}\
=\ c_{\pi_0(j)}^{\tau}(z,E,\eps)
\ee
where $\zeta_0$, $\theta_j(\zeta_0)$ and $\pi_0(j)$ are defined as above
and
are independent of $E\in\Delta$.
\end{lem}

\noindent
{\bf Remark:} Since $\Omega$ is finite,
it is straightforward to generalize the study of the analytic
continuations around one crossing point
to analytic continuations around several
crossing points. The loop $\zeta_0$ can be rewritten
as a concatenation of finitely many individual loops, 
each encircling only one point of $\Omega$.
The permutation $\pi_0$ is given by the composition of
associated permutations.
The factors $e^{-i\theta_j(\zeta_0)}$ in (\ref{theta}) are given by the
product of the factors associated with the individual loops.
The same is true for the factors
$\exp\left(\,-\,i\,\int_{\zeta_0}\,k_j^\tau(z,E)\,dz/\eps^2\,\right)$
in Lemma \ref{ECHCO}.

\vskip .5cm
We now describe how we use the above properties. 
The details may be found in \cite{joye}.

The idea is to integrate the integral equation corresponding to
(\ref{coef}) along paths that go above
(or below) one or several crossing points, and then to compare the result
with
the integration performed along the real axis.
As $z\ra -\infty $ in $\rho_\alpha$
these paths become parallel to the real axis so that the coefficients take
the same
asymptotic value ${c}_m^\tau(-\infty,E,\eps)$ along the real axis and the
integration paths.
Since the solutions to (\ref{adiab})
are analytic, the results of these integrations must agree as $\Re
z\ra\infty$.
Therefore,
(\ref{echco}) taken at $z=\infty$ yields the asymptotics of
$c_{\pi_0(j)}^{\tau}(\infty,E, \eps)$, provided we can control
$\widetilde{c}_j(z,E,\eps)$ in the complex plane.
We argue below that this can
be done in the so-called dissipative domains (See \cite{fed}, \cite{f1}), 
as proven in \cite{joye}.
We do not go into the details of these notions
because another result of
\cite{joye} will enable us to get sufficicient control on
$\widetilde{c}_j(z,E,\eps)$
in the avoided crossing situation, to which we restrict our attention.

We define
$$
{\Delta}_{jl}^{\tau\sigma}(x,E)\ =\
\int_0^x\,\left(\,k_j^\tau(y,E)\,-\,k_j^\sigma(y,E)\right)\,dy.
$$
By explicit computation, using formula (\ref{compa}) in (\ref{coef}),
we check that  (\ref{coef}) can
be extended to $\rho_\alpha\setminus\Omega$.
We next integrate by parts in (\ref{coef}), 
to see that (\ref{coef}) with $x_0=-\infty$ can be rewritten as
\bea\label{intpp}\hskip -12pt
  \widetilde{c}_m^\tau(z,E,\eps)&=&\delta_{jm}\,\delta_{\tau -}
  -i\eps^2\sum_{l, \sigma}\,
  \frac{\widetilde{a}_{ml}^{\tau
\sigma}(z,E)}{\widetilde{k}_m^{\tau}(z,E)-
  \widetilde{k}_l^\sigma(z,E)}\,
  e^{i\widetilde{\Delta}_{ml}^{\tau \sigma}(z,E)/\eps^2}\,
  \widetilde{c}_l^\sigma(z,E,\eps)
\\
  &&+\ i\eps^2\sum_{l,\sigma}\int_{-\infty}^z
  {\left(\frac{\partial}{\partial z'}\,
  \frac{\widetilde{a}_{ml}^{\tau \sigma}(z',E)}
{\widetilde{k}_m^{\tau}(z',E)-\widetilde{k}_l^\sigma(z',E)}\right)}\,
  e^{i\widetilde{\Delta}_{ml}^{\tau \sigma}(z',E)/\eps^2}\,
  \widetilde{c}_l^\sigma(z',E,\eps)\,dz'\nonumber
\\
  &&+i\eps^2\sum_{l,p,\sigma,\theta}\,\int_{-\infty}^z
  \frac{\widetilde{a}_{ml}^{\tau \sigma}(z',E)\,
      \widetilde{a}_{lp}^{\sigma \theta}(z',E)}
  {\widetilde{k}_m^{\tau}(z',E)-\widetilde{k}_l^\sigma(z',E)}\,
  e^{i\widetilde{\Delta}_{mp}^{\tau \theta}(z',E)/\eps^2}\,
  \widetilde{c}_p^{\theta}(z',E,\eps)\,dz',\nonumber
\eea
as long as the chosen path of integration does not meet $\Omega$.
Here, $\widetilde{\phantom{c}}$ denotes
the analytic continuation along the chosen path of
integration of the corresponding function defined originally on the real
axis.
This distinguishes $\widetilde{c}_m^\tau(\infty,E,\eps)$
from ${c}_m^\tau(\infty,E,\eps)$ computed along
the real axis as $x\ra\infty$.
These quantities may differ since the
integration path may pass above (or below) points of $\Omega$.

If the exponentials in (\ref{intpp}) are all uniformly bounded, as it is
the case
when the integration path coincides with the real axis, it is
straighforward
to get bounds of the type
\be\label{ifdis}
\tilde{c}_m^\tau(z,E,\eps)\ =\ \delta_{jm}\,\delta_{\tau -}\ +\ O_E(\eps^2).
\ee
In our context, all quantities depend on $E\in\Delta$. However,
by mimicking the proof of Proposition 4.1 of \cite{joye},
it is not difficult to check that the
estimate (\ref{ifdis}) is uniform for $E\in\Delta$.
For later purposes, we note that by differentiating (\ref{intpp}),
$\frac{\partial\phantom{E}}{\partial E}\widetilde{c}_m^\tau(z,E,\eps)$ is
uniformly bounded for $0<\eps<\eps_0$ and $E\in\Delta$ for any fixed $\eps_0$.

Again, as is well known, the existence of paths from $-\infty$ to $+\infty$
along which the exponentials do not blow up and
which pass above (or below) points in $\Omega$ is difficult to check in general.
It is linked to the global behavior of the Stokes lines of the
problem. See {\it e.g.}, \cite{fed}, \cite{f1}. 
This property goes under the name ``existence of
dissipative domains'' in \cite{joye}.

We avoid these complications by restricting attention to avoided crossing
situations where the existence has been proven \cite{joye}.

When dissipative domains exist, (\ref{echco}) and (\ref{ifdis}) imply
\be\label{wkb}
c_{\pi_0(j)}^{\tau}(\infty,E,\eps)\ =\
e^{-\,i\,\int_{\zeta_0}\,k^\tau_j(u,E)\,du/\eps^2}\
e^{-\,i\,\theta_j(\zeta_0)}\ (1+O_E(\eps^2)),
\ee
where the $O_E(\eps^2)$ estimate is uniform for $E\in\Delta$.
This is the main result of Proposition
4.1 in \cite{joye} in our context, under the assumption that dissipative
domains exist.
\\

\subsection{Avoided Crossings}

We now explore the avoided crossing situation, alluded to above,
that allows us to avoid considerations of the dissipative domains.
We now assume that $h(x)$ has the form $h(x,\delta)$ and 
satisfies Hypotheses {\bf H4} and {\bf H5}. 

We first check that the allowed pattern of avoided crossings 
for $\sigma(h(x,\delta))$
can be transfered to the eigenvalues of $H(x,E,\delta)$,
obtained from $h(x,\delta)$ by (\ref{adiabH}).\\

From the explicit formulae (\ref{sigH}), we see immediately that $x_c\in\R$
is
a real crossing point for the eigenvalues $e_j(x,0)$ and $e_l(x,0)$ of
$h(x,0)$
if and only if it is a real crossing point for the analytic eigenvalues
$k_j^\tau(x,E,0)$
and $k_l^\tau(x,E,0)$ of $H(x,E,0)$, for $\tau=+, -$. Moreover,
$$
\left.
\frac{\partial}{\partial x}\,(k_j^\tau(x,E)-k_l^\tau(x,E))\right|_{x=x_c}
\ =\
\tau\ \left.\frac{\frac{\partial}{\partial x}(e_l(x,0)-e_j(x,0))}
{\sqrt{2\,(E-e_j(x,0))}}\right|_{x=x_c},
$$
so that the real crossings for $H(x,E,0)$ are also generic, in the sense of 
(\ref{gc}).

\vskip 4mm
\noindent{\bf Remark:} Our assumption {\bf H7} on the parameter $E$  forbids
real
crossings between eigenvalues $k_j^\tau(x,E,0)$ and $k_l^\sigma(x,E,0)$,
with $\sigma\neq \tau$.\\

Regarding the ordering of the eigenvalues of $H(x,E,\delta)$, if those
of $h(x,\delta)$ are ordered as in (\ref{order}), we have
\be\label{reor}
-k_1(x,E,\delta)<\cdots <-k_m(x,E,\delta)<0< k_m(x,E,\delta)<\cdots
<k_1(x,E,\delta).
\ee
This means that the pattern of the crossings for the group of eigenvalues
$\{-k_j(x,E,0)\}_{j=1,\cdots, m}$ is the same as that for the eigenvalues
$\{e_j(x,0)\}_{j=1,\cdots, m}$. The pattern of the crossings for
the group $\{k_j(x,E,0)\}_{j=1,\cdots, m}$ is the reflection 
with respect to the horizontal axis
of the one for $\{e_j(x,0)\}_{j=1,\cdots, m}$.
Therefore, assumptions {\bf H5}, i), ii), iii)
are also satisfied for the eigenvalues of $H(x,E,0)$, for a relabeling from
$1$ to $2m$
of (\ref{reor}) with $\delta=0$, and $x$ close to $-\infty$.

To any given pattern of real crossings for the eigenvalues
$\{e_j(x,0)\}_{j=1,\cdots, m}$ of $h(x,0)$,
we associate a permutation $\pi $ of $ \{1,2,\cdots, m\}$ as follows.
Assume the eigenvalues are labeled in ascending order at $x=-\infty$,
as in property i) of {\bf H5}.
If $e_j(\infty,0)$ is the $k^{\mbox{\scriptsize th}}$ eigenvalue in
ascending order at $x=\infty$, the permutation
$\pi$ is defined by
\be\label{permut}
\pi(j)\ =\ k.
\ee
We call $\pi$ the permutation associated with $\sigma(h(x,0))$.
For small $\delta>0$, the real crossings turn into avoided crossings
on the real axis and conjugate complex crossing points appear close
to the real axis. Then $\pi$ corresponds to
the permutation $\pi_0$ (\ref{pi0}) for a loop $\zeta_0$
that surrounds all complex
crossing points in the upper half plane
that are associated with the avoided crossings.
\\

These properties of corresponding patterns of real 
crossings of the spectra of $h(x,\delta)$ and $H(x,E,0)$
immediately yield the following convenient relation between the
permutation $\pi$ associated with $\sigma(h(x,0))$ and the permutation
$\Pi$ associated with
$\sigma(H(x,E,0))$.
If we denote $\Pi$ by the obvious notation
$$
\Pi(j,\tau)\ =\ (k,\sigma),
$$
then we have
$$
\Pi(j,\tau)\ =\ (\pi(j),\tau),\qquad \mbox{for all}\qquad (j,\tau)\,\in\,
\{1,\cdots,m\}\times \{-,+\}.
$$

\vskip .5cm
We can now restate the main result of \cite{joye}
that describes the asymptotics of the
coefficients defined in (\ref{coef}),
adapted to our scattering framework for
incoming states entering from the left. (See (\ref{wkb0}).)
Intuitively, this result says that for small $\delta>0$,
dissipative domains exist,
provided the pattern of real crossings satisfies {\bf H5}.
Therefore, estimates of the type (\ref{wkb}) are true for certain indices
$j$ and $n$, determined by the permutation (\ref{permut}).
It is not difficult to see that
the permutation $\pi$ describes the successive exchanges of
eigenvalues one gets by following
a path in the complex plane that goes above or below all complex
crossing points of the eigenvalues $e_j(x,\delta)$
that are associated with the avoided crossings.

\begin{thm}\label{PERCO}
Let $h(x,\delta)$ satisfy {\bf H4} and {\bf H5}.
If $\delta>0$ is small enough, the
$\pi(j),j$ elements of the matrix ${\cal S}^{--}(E,\eps)$ in (\ref{bs}),
with $\pi(j)$ defined in (\ref{permut}), have small $\eps$ asymptotics for
all $j=1,\cdots, m$, given by
$$
{\cal S}^{--}_{\pi(j),j}(E,\eps)\ =\ 
\prod_{l=j}^{\pi(j)\mp1}e^{-i\theta_l(\zeta_l,\delta)}\,
e^{i\int_{\zeta_l}\,k_l(z,E,\delta)\,dz/\eps^2}
\left(1+O_{E, \delta}(\eps^2)\right),
\qquad\quad\pi(j)\ \left\{\,{ >j \atop <j}\right.
$$
where, for $\pi(j)>j$ (respectively $\pi(j)<j$), $\zeta_l$,
$l=j,\cdots,\pi(j)-1$ (resp. $l=j,\cdots,\pi(j)+1$), denotes a
negatively (resp. positively) oriented loop
based at the origin which encircles the complex crossing point $z_r$ only 
(resp. $\overline{z_r}$)
corresponding to the avoided crossing between $e_l(x,\delta)$ and
$e_{l+1}(x,\delta)$ (resp. $e_{l-1}(x,\delta)$) at $x_r$. The
$\int_{\zeta_l}k_l(z,E,\delta)\,dz$ denotes the integral along $\zeta_l$
of the
analytic continuation of $k_l(0,E,\delta)$,
and $\theta_l(\zeta_l)$
is the corresponding factor defined by (\ref{theta}).
\end{thm}

\noindent{\bf Remarks:} \\
i) Revisiting the proof of this theorem in \cite{joye}, we see that we can
choose $\delta>0$ small enough so that dissipative domains can be
constructed 
uniformly for $E\in\Delta$. This stems from the formula
$$
k_j(x,E,0)-k_l(x,E,0)\ =\
\frac{2(e_l(x,0)-e_j(x,0))}{k_j(x,E,0)+k_l(x,E,0)},
$$
whose  denominator can be controlled, close to the real axis,
uniformly for $E\in\Delta$.
\\
ii) When there is only one avoided crossing between level $j$ and $j+1$
stemming
from a real crossing at $x=x_0$, we have $j+1=\pi(j)$.
The theorem says
$$
{\cal S}^{--}_{(j+1),j}(E,\eps)\ =\ e^{-i\theta_j(\zeta_j,\delta)}\
e^{i\int_{\zeta_j}k_j(z,E,\delta)\,dz/\eps^2}\
\left(1+O_{E, \delta}(\eps^2)\right),
$$
where the negatively oriented loop $\zeta_j$ encircles the corresponding
complex
crossing point $z_0$, with $\Im z_0 >0$. Similarly, interchanging the
roles of $j$ and
$j+1$, it yields with $\bar{\zeta}_j$ the conjugate of the loop
${\zeta}_j$,
$$
{\cal S}^{--}_{j (j+1)}(E,\eps)\ =\ e^{-i\theta_j(\bar{\zeta}_j,\delta)}\
e^{i\int_{\bar{\zeta}_j}k_j(z,E,\delta)\,dz/\eps^2}
\left(1+O_{E, \delta}(\eps^2)\right),
$$
iii) Since the eigenvalues are continuous at the complex crossings, we
have
$$
\lim_{\delta\ra 0}\ 
\Im\,\int_{\zeta_j}\,k_j(z,E,\delta)\,dz\ =\ 0,\qquad\mbox{for all}\qquad
j=1,\cdots, p.
$$
It is shown in \cite{j} that
$$
\lim_{\delta\ra 0}\ \Im\,\theta_j(\zeta_j,\delta)\ =\ 0
\qquad\mbox{for all}\qquad j=1,\cdots, p.
$$
iv) The $O_{E, \delta}(\eps)$ errors in Theorem \ref{PERCO} depend on $\delta$,
but it should be possible to get estimates which are valid
as both $\eps$ and $\delta$ tend to zero, in the spirit of
\cite{j}, \cite{mn}, and \cite{r}.\\
v) This result shows that at least one off-diagonal element per column
of the $S$-matrix can be computed asymptotically.
However, it is often possible to get more elements by making use
of symmetries of the $S$-matrix. See \cite{joye} and \cite{jp2000}.\\

\vskip .4cm
In our avoided crossings context, transitions 
of the coefficients between states that
correspond to electronic levels that do not display
avoided crossings, {\it i.e.}, that are separated by a gap of order
1 as $\delta\ra 0$, are expected to be exponentially smaller than  the
transitions we control by means of Theorem \ref{PERCO},
as $\delta$ shrinks to zero.
Since the coefficients in the exponential decay rates given by the theorem
vanish in the limit $\delta\ra 0$, it is enough to show
that the decay rates of the exponentially small transitions
between well separated levels are independent of $\delta$.

That is the meaning of the following proposition, which
draws heavily upon \cite{jp3} and \cite{joye} and is proven in Section
\ref{technical}.

\begin{prop}\label{smaller} Let $F(x,\delta)$ be a $n\times n$ matrix
that satisfies {\bf H4}, except for the condition that $F(\cdot,\delta)$
be self-adjoint. Suppose its
eigenvalues $\{f_j(x,\delta)\}_{j=1,\cdots, m}$
that satisfy {\bf H5}.
Further assume that the eigenvalues
can be separated into two groups $\sigma_1(x,\delta)$ and
$\sigma_2(x,\delta)$ that display no avoided crossing,
{\it i.e.}, such that
$$
\inf_{\delta\geq 0\atop x\in \rho_\alpha\cup\{\pm\infty\}}
\mbox{\em dist}(\sigma_1(x,\delta),\,\sigma_2(x,\delta))\ \geq\  g\ >\ 0.
$$
Let $P(x,\delta)$ and $Q(x,\delta)=\un -P(x,\delta)$ be the projectors
onto the spectral subspaces corresponding to $\sigma_1(x,\delta)$ and
$\sigma_2(x,\delta)$ respectively, and let $U_\eps(x,x_0,\delta)$
be the evolution
operator corresponding to the equation
\be\label{evol}
i\,\eps^2\,\frac{d}{dx}\,U_\eps(x,x_0,\delta)\ =\ 
F(x,\delta)\ U_\eps(x,x_0,\delta),\qquad\mbox{with}\qquad
U_\eps(x_0,x_0,\delta)=\un .
\ee
Then, for any $\delta>0$, there exists $\eps_0(\delta)$,
$C(\delta)>0$ depending on $\delta$, and
$\Gamma >0$ independent of $\delta$, such that for all $\eps \leq
\eps_0(\delta)$,
$$
\lim_{x\ra\infty \atop x_0\ra -\infty}\
\|\,P(x,\delta)\,U_\eps(x,x_0,\delta)\,
Q(x_0,\delta)\,\|\ \leq\ C(\delta)\ e^{-\Gamma/\eps^2}.
$$
\end{prop}

\noindent{\bf Remark:} This proposition implies that reflections,
{\it i.e.},
the transitions from wave packets traveling to the right to wave
packets traveling to the left, on any electronic level, are exponentially
smaller than transitions associated with the avoided crossings in
which
the propagation direction is not changed. 
This is a consequence of Hypothesis {\bf H7} which implies that
complex crossings between $k_j^+$ and $k_l^-$, are far from the real axis
for any $j, l\in\{1,\cdots, m\}$.\\

Let us investigate more closely the analytic structure
of $k_j(z,E,\delta)$ in our avoided crossing regime characterized
by {\bf H4} and {\bf H5},
in order to deduce the
properties of the exponential decay rates 
$\Im \int_{\zeta_j}\,k_j(z,E,\delta)\,dz$. 
We do so for the $k_j$'s that correspond to
electronic eigenvalues $e_j(z,\delta)$ and 
$e_n(z,\delta)$ 
that experience only one 
avoided crossing, {\it i.e.}, we take $n=j\pm 1$. We can thus drop
the index $j$ in $\zeta_j$ in the notation.
We follow \cite{j} where a similar analysis is performed, sometimes
refering to results proven there.
The general case is dealt with by adding the corresponding
contributions stemming from each individual avoided crossing. \\

We can assume that the avoided crossing takes place at $x=0$, {\it i.e.},
$$
e_j(0,0)\ =\ e_n(0,0)\ \equiv\ e_c,
$$
where $e_c$ is the electronic eigenvalue at the crossing.
We also define the momentum $k_c(E)$ at the crossing point by
$$
k_c(E)\ =\ \sqrt{2\,(E-e_c)}
$$
and the quantity $\Gamma_0(\delta)$ by
\be\label{elephant}
\Gamma_0(\delta)\ =\ \left|\,\Im\int_\zeta\,e_j(z,\delta)\,dz\,\right| .
\ee
This quantity is the exponential decay rate given by the
Landau-Zener Formula for a (time dependent) adiabatic problem with
hamiltonian
$h(t,\delta)$. See \cite{j}. In Section \ref{technical} we prove 

\begin{lem}\label{dec} With the above notation, we have the following
as $\delta \ra 0$, uniformly for $E\in\Delta$, 
\bea\nonumber
&&\Im \int_\zeta\,\sqrt{2(E-e_j(z,\delta))}\,dz\ =\
\frac{\Gamma_0(\delta)}{k_c(E)}\ +\ O(\delta^3),\\ \nonumber
&&\frac{\partial}{\partial E}\ 
\Im \int_\zeta\,\sqrt{2(E-e_j(z,\delta))}\,dz\ =\
-\ \frac{\Gamma_0(\delta)}{k_c^3(E)}\ +\ O(\delta^3),
\qquad\qquad\mbox{and}\\ \nonumber
&&\frac{\partial^2}{\partial E^2}\
\Im \int_\zeta\,\sqrt{2(E-e_j(z,\delta))}\,dz\ =\
3\ \frac{\Gamma_0(\delta)}{k_c^5(E)}\ +\ O(\delta^3),
\eea
where
$0\ <\ \Gamma_0(\delta)\ = \ O(\delta^2)$.
\end{lem}

\vskip 2mm \noindent
{\bf Remarks:}\\
i) This implies that
$\Im \int_\zeta\,\sqrt{2(E-e_j(z,\delta))}\,dz$
is a positive, decreasing, convex function of $E$ on $\Delta$.
This remains true when the transition is mediated
by several avoided crossings.\\
ii) The first relation can be interpreted as saying that
in our Born--Oppenheimer context, the (time dependent adiabatic)
Landau-Zener decay
rate at fixed energy $E$ has to be modified in order to take into account
the
classical velocity $k_c(E)$ at the crossing.
\\
iii) More precise estimates will be derived below, further assuming 
{\bf H6}.
\\

\section{Exact Solutions to the Time--Dependent Schr\"odinger Equation}
\setcounter{equation}{0}

We now construct solutions to (\ref{schr}) by taking time--dependent
superpositions of the generalized eigenvectors $\Psi(x,E,\eps)$,
where $E\in\Delta$. 
%times a smooth energy density
%function with respect to $E\in\Delta$.
These superpositions depend on an energy density
$Q(E,\eps)$ that can be complex and may or may not depend on $\eps$. We always
assume that the following condition holds:\\

{\bf C0} : The density $Q(E,\eps)$ is $C^1$
as a function of $E\in\Delta$, for fixed $\eps>0$.\\

In this Section, the parameter $\delta>0$ is kept
fixed and plays no role. We therefore drop it from the notation
and work under hypotheses {\bf H1}, {\bf H2}, and {\bf H3}.\\

We define
\bea\hspace{-5mm}
\psi(x,t,\eps)&=&\int_\Delta\,Q(E,\eps)\,\Psi(x,E,\eps)\,e^{-itE/\eps^2}\,dE
\nonumber\\[2mm]
&=&\sum_{j=1,\cdots,m,\ \sigma=\pm}
\phi_j(x)\ \int_{\Delta}\,\frac{Q(E,\eps)}{\sqrt{2\,k_j(x,E)}}\,
c_j^\sigma(x,E,\eps)\,e^{-i\int_0^xk_j^\sigma(y,E)dy/\eps^2}\,
e^{-itE/\eps^2}\,dE
\nonumber\\[2mm]\label{sol}
&\equiv&\sum_{j=1,\cdots,m\ \sigma=\pm}\psi_j^\sigma(x,t,\eps).
\eea
Here $\psi_j^\sigma$ asymptotically describes the piece of the solution
that lives on the electronic state $\phi_j$ and propagates in the direction
characterized by $\sigma$.
Since the integrand is smooth and $\Delta$ is compact, 
$\psi(x,t,\eps)$ is an exact solution to the 
time--dependent Schr\"odinger equation (\ref{schr}). Note that this solution
is not necessarily normalized.

The following lemma, whose proof can be found in Section \ref{technical},
gives a bound that we use to understand the large $t$ behavior of
$\psi_j^\sigma(x,t,\eps)$. It is a simple corollary that
the state (\ref{sol}) belongs to $L^2(\R)$.

\begin{lem}\label{L2} Assume {\bf H1}, {\bf H2}, {\bf H3} and {\bf C0}.
Let 
$$
K_+\ =\ \sup_{j=1,\cdots,m\ E\in\Delta,\
  \sigma=\pm}\ k_j(\sigma\infty, E)
\ <\ \infty
$$
and 
$$
K_-\ =\ \inf_{j=1,\cdots,m\ E\in\Delta,\ \sigma=\pm}\ k_j(\sigma\infty, E)
\ >\ 0.
$$
Fix $\alpha\in(0,\,1)$. Then,
for either $t=0$, or
for any $x\neq 0$ and $t\neq 0$, such that 
$$
|x/t|\ >\ K_+/(1-\alpha)\qquad\mbox{or}\qquad
|x/t|\ <\ K_-/(1+\alpha),
$$
we have
$$
\left\|\,\psi_j^\sigma(x,t,\eps)\,\right\|\ \le\ 
C_{\eps}/|x|,\qquad\mbox{with}\ \ C_{\eps}\ \ \mbox{independent of}\quad t,
$$
where the estimate is in the norm on $\C^m$.
\end{lem}

\vskip 3mm
We now introduce freely propagating states
$\psi(x,t,\eps,\pm)\in L^2(\R,\,\C^m)$ that describe
the asymptotics of the solutions $\psi(x,t,\eps)$ as
$t\ra\pm\infty$. We use these asymptotic states
when we study the scattering matrix for (\ref{schr}).
We let
\bea\label{asst}\hspace{-5mm}\phantom{z}
&&\hspace{-1.3cm}\psi(x,t,\eps,\pm)\\[2.5mm]\hspace{-5mm}
&&\hspace{-1.3cm}=\quad\sum_{j=1,\cdots,m\ \sigma=\pm}\phi_j(x)\,\int_{\Delta}
\frac{Q(E,\eps)}{\sqrt{2k_j(\pm\infty,E)}}\ e^{-itE/\eps^2}\,
c_j^\sigma(\pm\infty,E,\eps)\,
e^{-i(xk_j^\sigma(\pm\infty,E)+\omega_j^\sigma(\pm\infty,E))/\eps^2}
\,dE\nonumber\\[4mm]\nonumber\hspace{-5mm}
&&\hspace{-1.3cm}=\quad\sum_{j=1,\cdots,m\ \sigma=\pm}\psi_j^\sigma(x,t,\eps,\pm)
\eea
These states are linear combinations of products of free
scalar wave packets in constant scalar potentials times
eigenvectors of the electronic Hamiltonian.
Their propagation is thus governed by the various channel
Hamiltonians.

\vskip 3mm
\begin{prop}\label{scat} Assume {\bf H1}, {\bf H2}, {\bf H3} and {\bf C0}.
In $L^2(\R)$ norm as $t\ra \pm\infty$, we have 
\be\label{prop}
\|\,\psi(x,t,\eps)\,-\,\psi(x,t,\eps,\pm)\,\|\ =\ O_\eps(1/|t|).
\ee
\end{prop}

\vskip 3mm \noindent
{\bf Remarks:}\\
i) The estimate (\ref{prop}) depends on $\eps$.\\
ii) By a change of variables, we immediately obtain the following corollary.

\vskip 3mm
\begin{cor}
The density of the component of the asymptotic momentum space wave function
on the $j^{\mbox{\scriptsize th}}$ electronic level
as $t \ra \pm\infty$ is
$$
\sigma\ \sqrt{\frac{k}{2}}\ Q(E(k),\eps)\ c_j^\sigma(\pm\infty,E(k),\eps)\
e^{-i\omega_j^\sigma(\pm\infty,E(k))/\eps^2}.
$$
Here $E(k)=k^2/2+e_j(\pm\infty)$
%$k=\sqrt{2(E-e_j(\pm \infty))}$ 
and $\sigma=-/+$ for waves
traveling in the positive/negative direction, respectively.
\end{cor}

\vskip 2mm \noindent
iii) Consider a solution $\psi(x,t,\eps)$ traveling
in the positive direction and
associated with the electronic eigenstate $\phi_{j}$ in the remote past.
It is characterized by 
$c_{k}^\sigma(-\infty,E,\eps)=\delta_{k,j}\,\delta_{\sigma,-}$,
and as $t\ra -\infty$, it is asymptotic to
\be\label{incomingwave}
\psi(x,t,\eps,-)\ =\ 
\phi_j(x)\ \int_{\Delta}\,\frac{Q(E,\eps)}{\sqrt{2k_j(-\infty,E)}}\,
e^{-itE/\eps^2}\,
e^{i(xk_j(-\infty,E)-\omega_j^-(-\infty,E))/\eps^2}\,dE.
\ee
As $t\ra +\infty$, the component of this state
that has made the transition from state $j$ to state $n$ is
asymptotic to the vector $\psi_n^-(x,t,\eps,+)$. 
It is given in terms of the matrix ${\cal S}$ by
\be\label{bon}
\phi_n(x)\ \int_{\Delta}\,\frac{Q(E,\eps)}{\sqrt{2k_n(+\infty,E)}}\,
e^{-itE/\eps^2}\,{\cal S}^{--}_{nj}(E,\eps)\,e^{+i(xk_n(+\infty,E)-
\omega_n^-(+\infty,E))/\eps^2}\,dE.
\ee
iv) Proposition \ref{scat} is proven in the Section \ref{technical}.

\section{Non--adiabatic Transition Asymptotics }
\setcounter{equation}{0}

\subsection{The Transition Integral}

From now on, we assume we are in the avoided crossing situation, but we
still do not make explicit the dependence in the variable $\delta>0$ 
in the notation.

Section 3 gave us the semiclassical
asymptotics of the elements of the $S$-matrix
${\cal S}(E,\eps)$. We now compute the small $\eps$ asymptotics
of the integrals that describe the asymptotic states
$\psi_j^\sigma(x,t,\eps,\pm)$
given by (\ref{asst}) as $|t|\ra\infty$, for the different channels.

We choose our energy density $Q(E,\eps)$ to be
more and more sharply peaked near a specific value $E_0\in \Delta\setminus\partial\Delta$
as $\eps\ra 0$. As a result, we obtain semiclassical Born-Oppenheimer states 
that are well localized in phase space.
This choice is physically reasonable, and it allows us to 
relate the quantum scattering process to classical quantities.

More precisely we consider, 
\be\label{howl} 
Q(E,\eps)\ =\ e^{-\,G(E)/\eps^2}\ e^{-\,i\,J(E)/\eps^2}\ P(E,\eps),
\ee
where \\[3mm]
{\bf C1} : The real-valued function $G\ge 0$ is in $C^3(\Delta)$,
and has a unique non-degenerate absolute minimum value of $0$
at $E_0$ in the interior of $\Delta$.
This  implies that 
$$
G(E)\ =\ g\,(E-E_0)^2/2\ +\ O(E-E_0)^3,\quad\mbox{ where }\quad g>0.
$$
{\bf C2} : The real-valued function $J$ is in $C^3(\Delta)$.\\[3mm]
{\bf C3} : The complex-valued
function $P(E,\eps)$ is in $C^1(\Delta)$ and satisfies
\be
\sup_{E\in\Delta \atop \eps \geq 0}\ 
\left|\,\frac{\partial^n}{\partial E^n}\,P(E,\eps)\,\right|\
\leq\ C_n,\qquad\mbox{for}\qquad n=0,\,1.\label{condd}
\ee

\vskip 2mm
\noindent
{\bf Remark:}\quad Typical interesting choices of $Q$ have 
$G\,=\,g\,(E-E_0)^2$, $J=0$, and
$P$ an $\eps$-dependent multiple of a smooth function
with at most polynomial growth in $(E-E_0)/\eps$.

\vskip 2mm
In our avoided crossing situation, we have already proved the following:\
A wave packet incoming from
the left in the remote past produces reflected waves ({\it i.e.},
components that travel to the left in the remote future)
that are exponentially smaller than the components that travel
to the right in the remote future.
We have also proved that the non-trivial transitions to electronic states
that are not involved in the avoided crossing are exponentially smaller than
those to electronic states that are involved in the avoided crossing.

Thus, the leading non-adiabatic transitions are described by the
asymptotics of those coefficients $c_l^\sigma(\pm\infty, E,\eps)$
that satisfy
\bea\label{choice}
c_k^\sigma(-\infty,E,\eps)&=&\delta_{j,k}\ \delta_{\sigma,-}\\[1mm]
\label{choice2}
c_n^-(+\infty,E,\eps)&=&e^{-i\theta_j(\zeta)}\ e^{i\int_{\zeta}k_j(z,E)dz/\eps^2}
\ (1+O_E(\eps^2)),
\eea
where $n=\pi(j)=j\pm 1$.
We recall that the error term $O_E(\eps^2)$ depends analytically on
the energy $E$ in a neighborhood of the compact set $\Delta$.
We have already noted in the comments after (\ref{ifdis}) that the term
$O_E(\eps^2)$ satisfies (\ref{condd}).

\vskip 2mm 
The form chosen for the energy densities should make it clear that Gaussian
wave packets will play a particular role in the asymptotic analysis of
(\ref{bon}). Therefore we use the specific 
notation introduced in (\ref{gcs}) for them.

Recall that a normalized free Gaussian state
propagating in the constant potential
$e(+\infty)$ is characterized by the classical quantities
\bea\nonumber
A_+(t)&=&A_+\,+\,i\,t\,B_+,\\ \nonumber
B_+(t)&=&B_+,\\ \nonumber
a_+(t)&=&a_+\,+\,\eta_+\,t,\\ \nonumber
\eta_+(t)&=&\eta_+,
\qquad\qquad\qquad\qquad\qquad\qquad\qquad\mbox{and}\\ \nonumber
S_+(t)&=&\int_0^t\ 
\left(\,\eta_+^2(s)/2\,-\, e(+\infty)\,\right)\,ds,
\eea
with $\Re (\overline{A_+}B_+)=1$ (see \cite{raise}).
The associated nuclear wave packet has the form
\bea\label{finought}
&&\hspace{-1.5cm}
e^{iS_+(t)/\eps^2}\ \ffi_0(A_+(t),B_+(t),\eps^2,a_+(t),\eta_+(t),x)\\[3mm]
&&\hspace{-1.5cm}=\quad
\frac{e^{it(\eta^2/2-e(\infty))/\eps^2}}{\pi^{1/4}\sqrt{\eps(A_++itB_+)}}
\ \exp\left\{-\frac{(x-(a_++\eta_+ t))^2 B_+}{2\eps^2 (A_++itB_+)}\right\}
\ \exp\left\{i\frac{\eta_+ (x-(a_++\eta_+ t))}{\eps^2}\right\}\nonumber.
\eea

We now have everything to state our main result:
\begin{thm}\label{mai}
Let $\psi(x,t,\eps)$ be a solution of the  Schr\"odinger
equation (\ref{schr}) with electronic Hamiltonian $h(x,\delta)$
that satisfies hypotheses {\bf H4}, {\bf H5}, {\bf H7}. Assume that
the solution is characterized asymptotically in the past by
$$ 
\lim_{t\ra -\infty}\ \|\,\psi(x,t,\eps)\,-\,\psi^-_j(x,t,\eps,-)\,\|\ =\ 0,
$$
with
$$
\psi^-_j(x,t,\eps,-)\ =\ \phi_j(x)\
\int_{\Delta}\ \frac{Q(E,\eps)}{\sqrt{2k_j(-\infty,E)}}\ 
e^{-itE/\eps^2}\,
e^{i(xk_j(-\infty,E)-\omega_j^-(-\infty,E))/\eps^2}\,dE,
$$
where the energy density is supported on the interval $\Delta$, and
$$ 
Q(E,\eps)=\ e^{-\,G(E)/\eps^2}\ e^{-\,i\,J(E)/\eps^2}\ P(E,\eps)
$$
satisfies {\bf C1}, {\bf C2}, and {\bf C3}. Let $n=\pi(j)$ be given by
(\ref{permut}), and let 
\bea\label{al}
\alpha(E)&=&G(E)\,+\, \Im (\int_{\zeta}\,k_j(z,E)\,dz),\\%[-1mm]
\label{ka}
\kappa(E)&=&J(E)\,-\, \Re(\int_{\zeta}\,k_j(z,E)\,dz)\,+\,\omega_n^-(\infty,E).
\eea
Assume $E^*$ is the unique absolute minimum of $\alpha(\cdot)$
in {\em Int}\,$\Delta$.

Then, there exist $\delta_0>0$, $p>0$ arbitrarily close to 3, and a function
$\eps_0:(0,\delta_0)\ra \R^+$, such that for all
$\delta <\delta_0$ and $\eps< \eps_0(\delta)$, 
the following asymptotics hold as $t\ra \infty$:
\bea\hspace{-1cm}
\psi_n^-(x,t,\eps, +)&=&\nonumber
\phi_n(x)\ 
\frac{e^{-\,i\,\theta_j(\zeta)}\
\eps^{3/2}\ \pi^{3/4}}{\left(\frac{d^2}{dk^2}\alpha(E(k))|_{k^*}\right)^{1/4}}
\ e^{iS_+(t)/\eps^2}\ \ffi_0(A_+(t),B_+(t),\eps^2,a_+(t),\eta_+(t),x)
\\[3mm]
&&\hspace{-3mm}\times\ 
P(E^*,\eps) \sqrt{k^*} e^{-\alpha(E^*)/\eps^2}\
e^{-i(\kappa(E^*)-{k^*}^2\kappa'(E^*))/\eps^2}+
O(e^{-\alpha(E^*)/\eps^2}\eps^{p})  
+O_\eps
\left(1/t\right),\nonumber
\eea
where $\ffi_0$ is parametrized by
\bea\nonumber
&&\eta_+=k^*=\sqrt{2(E^*-e_n(\infty))},\quad\
a_+=k^*\,\kappa'(E^*),\quad\
B_+=\frac{1}{\sqrt{\frac{d^2}{dk^2}\alpha(E(k))|_{k^*}}},
\\[2mm] \label{ident}
& & A_+=\left(\,\frac{d^2}{dk^2}\alpha(E(k))|_{k^*}+i
\frac{d^2}{dk^2}\kappa(E(k))|_{k^*}\,\right)
\left/\sqrt{\,\frac{d^2}{dk^2}\alpha(E(k))|_{k^*}\,}\right. 
\qquad\mbox{and}\\\nonumber
& & S_+(t)=t({k^*}^2/2 - e_n(\infty)).\eea
All error terms are estimated in the $L^2(\R)$ norm, and
the estimate $O(e^{-\alpha(E^*)/\eps^2}\eps^{p})$
is uniform in $t$, whereas $O_\eps(1/t)$ may depend on $\eps$.
\end{thm}
\vskip 3mm
\noindent
{\bf Remarks:}\\
i)\quad All quantities computed from the electronic Hamiltonian $h(x,\delta)$ 
depend on $\delta$, even though that dependencs is not
specified in the notation.\\ 
ii)\quad The function $\alpha$ has a unique absolute minimum if 
$|\Delta|$ and $\delta$ are small enough. See Proposition \ref{51}. However,
in the case of several absolute minima, one simply adds the contributions 
associated with each of them. \\
iii)\quad The transitions to states that travel to the left
in the future are excluded
from our analysis because of the lack of uniformity in $E$ in the
semiclassical asymptotics of the relevant elements of the matrix
${\cal S}(E,\eps)$. At the price of some more technicalities,
it should also be possible to accommodate this situation by
our methods.\\
iv)\quad When several avoided-crossings are taken into account and
meet the requirements of Theorem \ref{PERCO}, $c_n^-(\infty,E,\eps)$
with $n=\pi(j)$ is given
by a product of exponentials of the same form as those in (\ref{choice2}).
The analysis of this situation is essentially identical to the
single avoided crossing situation, {\it mutatis mutandis}.\\
v)\quad Further properties of
$\psi_n(x,t,\eps,+)$ are given below. In particular, the characteristics
of the average momentum $k^*$ and its behavior as a function of $\delta$
are detailed in Section 5.2. The energy densities corresponding to specific 
incoming states are studied in Section 6.\\
vi)\quad The asymptotics of the incoming wave
with the electrons in the state $\phi_j$ in the remote past
are described by the same integral, with the replacements
\be\label{changes}
\left\{\ \begin{array}{l}\gamma_j\ \mapsto\  0, \\[2mm]
\omega_n^-(\infty,\cdot)\ \mapsto\ \omega_j^-(-\infty,\cdot),\\[3mm]
\sqrt{2(E-e_n(\infty))}\ \mapsto\ \sqrt{2(E-e_j(-\infty))},\\[2mm]
\theta_j\mapsto 0.\end{array}\right.
\ee

\vskip 4mm\noindent
{\bf Proof of Theorem \ref{mai}:}\\
Apart from the $E$-independent factor given by
$$
\frac{\phi_n(x)}{\sqrt{2}}\ e^{-\,i\,\theta_j(\zeta)},
$$
the asymptotics of (\ref{bon}) are determined by the integral
\bea\nonumber
T(\eps,x,t)&=&
\int_{\Delta}\,\frac{\tilde{P}(E,\eps)}{(2(E-e_n(\infty)))^{1/4}}
\\[2mm]&&\quad\times\quad
e^{-G(E)/\eps^2}\ e^{-i(tE+J(E))/\eps^2}\ e^{i\gamma_j(E)/\eps^2}\
e^{i(x\sqrt{2(E-e_n(\infty))}-\omega_n^-(\infty,E))/\eps^2}\ dE,\nonumber
\eea
where $\tilde{P}(E,\eps)=P(E,\eps)\,(1+O_E(\eps^2))$ satisfies
(\ref{condd}),
\bea\nonumber
\gamma_j(E)&=&\int_{\zeta}\,k_j(z,E)\,dz,\qquad\mbox{and}\\[2mm]
\nonumber\omega_n^-(\infty,E)&=&-\int_0^\infty\ 
\left(\,\sqrt{2(E-e_n(y))}\,-\,\sqrt{2(E-e_n(\infty))}\,\right)\ dy.
\eea
The $(1+O_E(\eps^2))$ factor in  $\tilde{P}(E,\eps)$ comes from Theorem \ref{PERCO}.
Recall that $\gamma_j$ and $\omega_j^+(\infty,\cdot)$  are analytic in a complex neighborhood
of $\Delta$, and that $\Im \gamma_j(E)$ is a positive,
decreasing, convex function of $E$, for $\delta$ sufficiently small.

\vskip 3mm
In terms of the  $C^3$ functions (\ref{al}) and (\ref{ka})
we can write $T(\eps,x,t)$ as
$$
T(\eps, x, t)\ =\ \int_{\Delta}\
\frac{\tilde{P}(E,\eps)}{\left(2(E-e(\infty)\right)^{1/4}}
\ e^{-\alpha(E)/\eps^2}\ e^{-i(tE+\kappa(E))/\eps^2}\
e^{i(x\sqrt{2(E-e(\infty))}/\eps^2}\,dE,
$$
where we have dropped the index in the asymptotic eigenvalue
$e(\infty)=e_n(\infty)$.
In Section \ref{technical} we analyze the small $\eps$ asymptotics of
$T$ essentially by Laplace's method. The result is

\begin{lem}\label{icet}
Let
$\ds k(E)\,=\,\sqrt{2(E-e(\infty))}$,
or equivalently,
$\ds E(k)\,=\,\frac{k^2}{2}+e(\infty)$,
and assume that $\alpha(\cdot)$ has a unique absolute minimum $E^*$.
For sufficiently small $\delta$, 
this minimum is non-degenerate and satisfies
$E^*\in\mbox{Int}\ \Delta$.
With $k^*=k(E^*)$, there exists $p>0$ arbitrarily close to 3,
such that as $\eps\ra 0$,
\bea\label{tgauss}
&&\hspace{-1.5cm}
T(\eps, x, t)\ =\ 
\frac{k^*\frac{d^2}{dk^2}\alpha(E(k))|_{k^*}+i(x+{k^*}^3\kappa''(E^*))}
{(\frac{d^2}{dk^2}\alpha(E(k))|_{k^*}
+i(t+\frac{d^2}{dk^2}\kappa(E(k))|_{k^*}))^{3/2}}\\[4mm]
&&\qquad\times\quad\eps\ \sqrt{2\pi}\ \frac{P(E^*,\eps)}{\sqrt{k^*}}
\ e^{-\alpha(E^*)/\eps^2}\ 
\exp\left\{-i\frac{(tE^*+\kappa(E^*)-xk^*)}{\eps^2}\right\}
\nonumber\\[3mm]
&&\qquad\times\quad
\exp\,\left\{-\,\frac{(x-k^*(t+\kappa'(E^*)))^2}
{2\eps^2\left(\frac{d^2}{dk^2}\alpha(E(k))|_{k^*}+i(t+
\frac{d^2}{dk^2}\kappa(E(k))|_{k^*}\right)}\right\}\
+\ O(e^{-\alpha(E^*)/\eps^2}\eps^{p}),\nonumber
\eea
where the error estimate is in the $L^2(\R)$ norm, uniformly in $t$.
\end{lem}

\noindent
{\bf Remarks:}\\
i)\quad If there are several absolute minima, one simply adds their
contributions to get the asymptotics of $T$.\\
ii)\quad If $T$ is associated with the incoming wave as
$t\ra-\infty$, the formula holds with $E_0$ in place of $E^*$, $k_0=\sqrt{2(E_0-e(-\infty))}$
in place of  $k^*$, and the changes (\ref{changes}).\\
iii)\quad If $P$ satisfies {\bf C3} and $P(E^*,\eps)=O(\eps^d)$ for some
$d\ge \frac 32$, then
the above analysis yields no information.

\vskip .5cm
To relate the integral $T$ to standard Born--Oppenheimer
states involving normalized free Gaussian states,
we must identify (\ref{tgauss})
with (\ref{finought}), making use of (\ref{ident}), and taking care of the
$x$ and $t$ dependence in the non-Gaussian part of (\ref{tgauss}).
That is the content of 
the next lemma which completes the proof of Theorem \ref{mai}.
\\

With the identifications (\ref{ident}), we have

\begin{lem}\label{outgaus} For small $\eps$ and $0<p<3$, we have
\bea\nonumber\hspace{-7mm}
T(\eps,x,t)&=&
\eps^{3/2}\ 2^{1/2}\ \pi^{3/4}\ 
\left(\frac{d^2}{dk^2}\alpha(E(k))|_{k^*}\right)^{-1/4}\,
\frac{P(E^*,\eps)}{\sqrt{k^*}}
\ e^{-\alpha(E^*)/\eps^2}\\[3mm]
&&\times\quad
e^{-i(\kappa(E^*)-{k^*}^2\kappa'(E^*))/\eps^2}\
\left(\,
\frac{k^*\frac{d^2}{dk^2}\alpha(E(k))|_{k^*}
+i(x + {k^*}^3\kappa''(E^*))}
{\frac{d^2}{dk^2}\alpha(E(k))|_{k^*}+
i(t+\frac{d^2}{dk^2}\kappa(E(k))|_{k^*})}
\,\right)
\nonumber \\[3mm] \nonumber
&&\times\quad
%\
e^{iS_+(t)/\eps^2}\ffi_0(A_+(t),B_+(t),\eps^2,a_+(t),\eta_+(t),x)
\quad +\quad
O(e^{-\alpha(E^*)/\eps^2}\eps^{p}),
\eea
where the error is estimated in the $L^2(\R)$ norm, uniformly in $t$.\\

Furthermore, in the $L^2$ norm, for small $\eps$  and  large $|t|$,
we have
\bea\nonumber
T(\eps,x,t)&=&\eps^{3/2}\ 2^{1/2}\ \pi^{3/4}\ {P(E^*,\eps)}\ {\sqrt{k^*}}\ e^{-\alpha(E^*)/\eps^2}\
e^{-i(\kappa(E^*)-{k^*}^2\kappa'(E^*))/\eps^2}\\[3mm]
&&\times\quad e^{iS_+(t)/\eps^2}\
\ffi_0(A_+(t),B_+(t),\eps^2,a_+(t),\eta_+(t),x)\
\left(\frac{d^2}{dk^2}\alpha(E(k))|_{k^*}\right)^{-1/4}
\nonumber\\[3mm]\nonumber
&&+\quad O(e^{-\alpha(E^*)/\eps^2}\eps^{p})\quad +\quad O(\eps^{3/2}/|t|),
\eea
where the first error term is uniform in $t$.
\end{lem}

\noindent
{\bf Remarks:} \\
i)\quad We note that the quantities $\alpha(\cdot)$, $k^*$, and
$B_+$ depend only on the index $j$, while
$\kappa(\cdot)$, and hence,  $A_+$ depend on both $j$ and $n$.
\\
ii)\quad More detailed computations are carried out in the
next section, which is devoted to specific incoming states.

\hfill \ep

\subsection{Energy and Momentum Shifts}

When there is a single avoided crossing, we can be more precise
about the energy and momentum shifts revealed by our general analysis.

For the rest of this section, we assume $h(x,\delta)$ satisfies
Hypothesis {\bf H6}.

Under this hypothesis, it is known \cite{j} that the decay rate in
the Landau--Zener formula (\ref{elephant}) has the form
$$
\Gamma_0(\delta)\ =\
\delta^2\,\frac{\pi}{4}\,
\left(\frac{b^2}{a}-\frac{c^2}{a^3}\right)\,+\,O(\delta^3)
\ \equiv\ \delta^2\,D\,+\,O(\delta^3),
$$
and that
$$
\Im\theta_j(\zeta_j,\delta)\ =\ 0(\delta).
$$
We use these formulas to get more information on $E^*$,
the typical energy of the outgoing wave packet,
that is determined by
the relation
\be\label{defes}
\alpha'(E^*)\ =\ G'(E^*)\,+\,\Im \gamma_j'(E^*)\ =\ 0,
\ee
where the primes denote derivatives with respect to $E$.\\

In the next proposition, we consider two cases:

In the first case, we choose the exponent $G(E)$ in the energy density
to be independent of $\delta$.
This yields less interesting momentum and energy shifts since they
vanish to leading order in $\delta$ as $\delta\ra 0$, in keeping
with \cite{hagjoy1}.

In the second case, we choose $G(E)$ to depend on $\delta$ in such a way
that the incoming wave packet contains a sufficiently wide spectrum
of energies as $\delta\ra 0$.
This implies non-trivial behavior of the relevant
quantities to leading order in $\delta$.
For obvious reasons, we restore $\delta$ in the notation
of this discussion.\\

\begin{prop}\label{51}
Let
\bea\nonumber
&&G(E)\ =\ g(E-E_0)^2/2\,+\,O(E-E_0)^3,\\[1mm]
&&\Im\gamma_j(E,\delta)\ =\ 
\frac{\Gamma_0(\delta)}{k_c(E)}\,+\,O(\delta^3)\ =\
\frac{D\,\delta^2}{k_c(E)}\,+\,O(\delta^3),
\qquad\mbox{and}\nonumber\\[1mm]
&&\alpha(E,\delta)\ =\ G(E)\,+\,\Im\gamma_j(E,\delta),\nonumber
\eea
as above.\\
{\bf i)}\quad 
Assume $G$ is independent of $\delta$. Then, for $E^*$ defined by
(\ref{defes}), we have
$$
E^*(\delta)\ =\ E_0\,+\,\frac{\Gamma_0(\delta)}{g\,k_c^3(E_0)}
\ +\ O(\delta^3),
$$
as $\delta \ra 0$. In this case,
\bea\nonumber
\alpha(E^*(\delta))&=&
\frac{\Gamma_0(\delta)}{k_c(E_0)}\,+\,O(\delta^3)\ >\ \alpha(E_0),
\qquad\mbox{and}\\[1mm]
\alpha''(E^*(\delta))&=&g\,+\,O(\delta^2).\nonumber
\eea
If $G(E)\ =\ g(E-E_0)^2/2\ +\ g_1(E-E_0)^3/6\ +\ O(E-E_0)^4$, then
$$
\alpha''(E^*(\delta))\ =\
g\,+\,g_1\,\frac{\Gamma_0(\delta)}{g\,k_c^3(E_0)}\,+\,
3\,\frac{\Gamma_0(\delta)}{k_c^5(E_0)}\,+\,O(\delta^3).
$$
{\bf ii)}\quad
Assume $G(E,\delta)=L(\delta(E-E_0))$, for some function $L$,
such that
$$
G(E,\delta)\ =\ g_0\,\delta^2\,(E-E_0)^2/2\,+\,O(\delta^3),
$$
for some $g_0>0$, uniformly for $E\in\Delta$. Then
$$
E^*(\delta)\ =\ E_1\,+\,O(\delta),
$$
where $0<E_1=E_1(D/g_0)$ is the unique solution to the equation
$$
(E_1-E_0)\ =\ D/(g_0k_c^3(E_1)),
$$
and is independent of $\delta$. In this case,
\bea\nonumber
\alpha(E^*(\delta))&=&
\delta^2\,\left(\,\frac{D}{k_c(E_1)}\,+\,g_0(E_1-E_0)^2/2\,\right)
\,+\,O(\delta^3)\\ \nonumber
&=&\delta^2\,
\left(\,D^{2/3}g_0^{1/3}(E_1-E_0)^{1/3}+g_0(E_1-E_0)^2/2\right)
+O(\delta^3)\\
&>&\alpha(E_0),\qquad\qquad\qquad\mbox{and}\nonumber\\[4mm] \nonumber
\alpha''(E^*(\delta))&=&\delta^2\,g_0\,+\,
3\,\frac{\Gamma_0(\delta)}{k_c^5(E_0)}\,+\,O(\delta^3).
\eea
\end{prop}

\vskip 4mm
\noindent
{\bf Proof:}\quad
Both statements follow from application of the
Implicit Function Theorem
and the observation that $\alpha$ is a strictly 
convex function of $E$ on $\Delta$.
\hfill\ep\\

\noindent
{\bf Remarks:}\\
a)\quad The first result shows no effect to leading
order in $\delta$ in the exponential decay rate of transition probability.
The value of $E^*(\delta)$ and the width of the outgoing wave packet
can be computed. Their variations with respect to the corresponding
quantities in the incoming wave packet are of order $\delta^2$,
and hence, are rather small.\\
b)\quad In case ii) of the proposition, the equation that determines $E^*$
can be rewritten as the quintic equation
$$
k_c^5(E)\,-\,k_c^3(E)\,k_c^2(E_0)\,-\,2D/g_0\ =\ 0.
$$
c)\quad In the case ii),
the variation of exponential decay rate in the transition
probability is given by
\bea
&&\alpha(E^*(\delta))-\alpha(E_0)\ =\
\delta^2\left(D\left(\frac{1}{k_c(E_1)}-\frac{1}{k_c(E_0)}\right)
\,+\,g_0(E_1-E_0)^2/2\right)\ 
+\ O(\delta^3)\nonumber\\ \nonumber
&&=\ \frac{\delta^2\ D}{2\,g_0\,k_c(E_1)^6\,k_c(E_0)}\ 
\left(\,2\,g_0\,k_c(E_1)^5\,(k_c(E_0)-k_c(E_1))\,+\,D\,k_c(E_0)\,\right)\
+\ O(\delta^3).
\eea
d)\quad The results above hold provided one knows $E^*$ is
the unique absolute minimum of $G$ in the set $\Delta$,
which is generically true. Again,
if there are several minima, one simply adds the corresponding contributions.

\vskip 4mm
\section{Energy Densities and Transitions when the Incoming
State has the form $\ffi_m$}
\setcounter{equation}{0}

In this section we study the special case in which the incoming state is
asymptotically in the past on electronic level $j$ with the nuclear wave function
given by one of the functions $\ffi_m$. For simplicity, we restrict
attention to wave packets that are incoming from the left.

In the simplest situation, the incoming wave packet is asmyptotic to
\be\label{gaussin}
e^{i(\eta_-^2/2-e_j(-\infty))t/\eps}\
\ffi_0(A_-+itB_-,B_-,\eps^2,a_-+\eta_-t,\eta_-,x)\ \phi_j(x),
\ee
as $t\rightarrow -\infty$. Here we choose $\eta_->0$ and the set
$\Delta$, so that $\eta_-^2/2+e_j(-\infty)$ is in the interior of $\Delta$,
and that the minimum of $\Delta$ lies strictly above the spectrum 
of $h(x)$ for all $x$.

We choose a smooth cut--off function $F(E)$ whose support is a subset of
the interior of
$\Delta$, which takes the value $1$ on an interval whose interior contains
$\eta_-^2/2+e_j(-\infty)$,
and whose length is almost as large as that of $\Delta$.

From our assumptions on $\Delta$, there is a one-to-one correspondence between 
$E\in\Delta$ and positive $k$, such that $k^2/2+e_j(-\infty)=E$.
For $E\in\Delta$, we make the change of variables
from $k$ to $E$ at $t=0$ in the (rescaled) Fourier transform of the
Gaussian in (\ref{gaussin}) (see \cite{raise}). Taking into account 
the normalization 
(\ref{ffi}) of the generalized eigenvectors,
this leads to the energy density 
\bea\label{Qin}
Q(E,\eps) &=&
\frac{ F(E)\ e^{i\,\omega_j^-(E,-\infty)/\eps^2}}{\eps\ \sqrt{\pi\,k(E)}}\quad
\ffi_0\left( B_-,A_-,\eps^2,\eta_-,-a_-,k(E)\right)\\ \nonumber
&=&\frac{ F(E)\ e^{i\,\omega_j^-(E,-\infty)/\eps^2}}{\eps\ \sqrt{\pi\, 
\sqrt{2(E-e_j(-\infty))} }}\quad
\ffi_0\left( B_-,A_-,\eps^2,\eta_-,-a_-,\sqrt{2(E-e_j(-\infty))}\right)
\eea
that we use in (\ref{asst}).

Since $\eta_-^2/2+e_j(-\infty)$ is in the interior of the set where
$F(E)=1$, 
the wave functions (\ref{gaussin}) and $\psi(x,t,\eps,-)$ defined by
(\ref{asst}) with the energy density defined by (\ref{Qin}) differ 
in $L^2(\R)$ norm by
an $O(e^{-C/\eps^2})$ error.
To be sure that this error is smaller than the non--adiabatic effect
we are studying, we assume any one of the following conditions: 
\\
1.\quad Take the avoided crossing gap $\delta$ to be small enough 
that the non--adiabatic effect is larger than the error we make here.
\\
2.\quad Choose $|B_-|$ to be sufficiently small. That 
increases the value of $C$ in this error estimate.
\\
3.\quad Fix the minimum of $\Delta$, but then
choose $\eta_-$ large enough so that the cut off is farther out in the
tail of the Gaussian in momentum space. This also makes the non-adiabatic effect
larger since $\eta_-$ is larger.

\vskip 5mm
With $Q(E,\eps)$ chosen by (\ref{Qin}), we have
in the notation of (\ref{howl}), 
\bea\label{Gin}
G(E)&=&(\Re (A_-/B_-))\quad
\frac{(\sqrt{2\,(E-e_j(-\infty))}\,-\,\eta_-)^2}2\\[3mm]
\nonumber &=&
|B_-|^{-2}\quad\frac{(\sqrt{2\,(E-e_j(-\infty))}\,-\,\eta_-)^2}2,\\[4mm]
\label{Jin}
J(E)&=&(\Im (A_-/B_-))\quad \frac{(\sqrt{2(E-e_j(-\infty))}-\eta_-)^2}2\\[3mm]
\nonumber
&&\qquad\qquad\qquad\qquad\quad  +\quad
a_-\ \,(\sqrt{2(E-e_j(-\infty))}-\eta_-)\
-\ \omega_j^-(E,-\infty),\\[4mm] \label{Pin_0}
P(E,\eps)&=&\pi^{-3/4}\ \eps^{-3/2}\ B_-^{-1/2}\ 
(2(E-e_j(-\infty)))^{-1/4}\ F(E).
\eea 
Also, conditions {\bf C1},
and {\bf C2} are satisfied, and provided we remove the trivial normalization
factor of $\eps^{-3/2}$ from $P(E,\eps)$,
then condition {\bf C3} is also satisfied. 

We already know that asymptotically in the past, the interacting wave function
determined by (\ref{Qin}) agrees with (\ref{gaussin}) up to
an $O(e^{-C/\eps^2})$ error, and we observe that the density
$Q(E,\eps)$ is sharply peaked around the energy
$$
E_0\ =\ \frac{\eta_-^2}{2}\,+\,e_j(-\infty) \ \ \ \mbox{ corresponding to } \ \ \
 \eta_-\ =\ \sqrt{2(E_0-e_j(-\infty))}.
$$
Thus from (\ref{Gin}), we see that 
$$
G(E)\ =\ \frac{(E-E_0)^2}{2\,(\eta_-\,|B_-|)^{2}}\ +\ O((E-E_0)^3), \ \ \
\mbox{{\it i.e.},} \ \ \
g\ =\ \frac{1}{(\eta_-\,|B_-|)^{2}}.
$$

\vskip 4mm
We are not particularly interested in the main component of the wave
function for large time that has not made a non--adiabatic transition.
However, by a similar analysis, it could be determined by our techniques.
Of course, it is what one would expect from the standard time--dependent
Born--Oppenheimer approximation.

Our focus is on the dominant non--adiabatic component, which is determined
to leading order in $\eps$ by Theorem \ref{mai}. 
From the above calculations and Theorem \ref{mai}, we immediately get
our main result for Gaussian incoming states:

\begin{thm}\label{at_last_0}  Assume Hypotheses {\bf H4}, {\bf H5}, and
{\bf H7},
and assume $\Delta$, $A_-$, $B_-$, $a_-$, $\eta_-$, $\delta$,
and the levels $j$ and $n$ have been
chosen to satisfy the requirements above.
Let $\Psi(x,\eps,t)$ be the solution to the Schr\"odinger
equation that is asymptotic as $t\rightarrow -\infty$ to
$$
e^{i(\eta_-^2/2-e_j(-\infty))t/\eps^2}\ \,
\ffi_0(A_-+itB_-,B_-,\eps^2,a_-+\eta_-t,\eta_-,x)\ \,\phi_j(x).
$$
The leading non--adiabatic component of $\Psi(x,\eps,t)$ as
$t\rightarrow\infty$ and $\eps\ra 0$ in $L^2$ norm is on electronic level 
$\phi_n(x)$ and is given by
$$
{\cal A}_{nj}(\eps)\ \,
e^{i(\eta_+^2/2-e_n(\infty))t/\eps^2}\ \,
\ffi_0(A_++itB_+,B_+,\eps^2,a_++\eta_+t,\eta_+,x)\ \,\phi_n(x),
$$
where the values of $A_+$, $B_+$, $a_+$, $\eta_+=k^*$ are those
given by (\ref{ident}) as in Theorem \ref{mai}.
The amplitude for making this transition
from level $j$ to level $n$
is given by
$$
{\cal A}_{nj}(\eps)\
=\ e^{-i\theta_j(\zeta)}\ e^{-\alpha(E^*)/\eps^2}\
\sqrt{\frac{B_+}{B_-}}\ e^{-i(\kappa(E^*)-k^*a_+)/\eps^2}.
$$
In particular
\bea
B_+&=&((G''(E^*)+\Im \gamma_j''(E^*))\,{k^*}^2)^{-1/2}\ =\
\left(\,\frac{\eta_-}{|B_-|^2\,k^*}\
+\ \Im \gamma_j''(E^*)\,{k^*}^2\,\right)^{-1/2}\nonumber\\ \label{b+}
&=&\frac{|B_-|}{\sqrt{\frac{\eta_-}{k^*}+\Im \gamma_j''(E^*){|B_-|k^*}^2}}
\eea
\end{thm}
{\bf Remark:}\quad
Depending on the relative size of $|B_-|$ with respect to $\delta$,
we can apply Proposition \ref{51} to further characterize ${\cal A}_{nj}$.

\vskip 7mm
We now turn our attention to the situation where the incoming
nuclear wave packet is in the state $\ffi_m$. The only change
from the situation just considered is that we must replace
the function $P(E,\eps)$ in (\ref{Pin_0}) by
\bea\nonumber
P(E,\eps)&=&(-i)^m\,2^{-m/2}\,(m!)^{-1/2}\,
\pi^{-3/4}\,\eps^{-3/2}\,(2(E-e_j(-\infty)))^{-1/4}
B_-^{-(m+1)/2}\,(\overline{B_-})^{m/2}\\[3mm]
&&\times\quad
H_m\left(\frac{\sqrt{2(E-e_j(-\infty))}\,-\,\eta_-}{\eps\,|B|}\right)
\,F(E).\label{Pin_m}
\eea
Again, this satisfies Condition {\bf C3} if we take out the trivial
factor of $\eps^{-(m+3/2)}$.

\vskip 2mm
\noindent

\begin{thm}\label{at_last_m}  Assume the Hypotheses of Theorem \ref{at_last_0}.
Let $\Psi(x,\eps,t)$ be the solution to the Schr\"odinger
equation that is asymptotic as $t\rightarrow -\infty$ to
$$
e^{i(\eta_-^2/2-e_j(-\infty))t/\eps^2}\ \,
\ffi_m(A_-+itB_-,B_-,\eps^2,a_-+\eta_-t,\eta_-,x)\ \,\phi_j(x).
$$
The leading non--adiabatic component of $\Psi(x,\eps,t)$ as
$t\rightarrow\infty$ and $\eps\ra 0$ in $L^2$ norm is on electronic level 
$\phi_n(x)$, and is given  by
$$
{\cal A}_{nj}^{(m)}(\eps)\ \,
e^{i(\eta_+^2/2-e_n(\infty))t/\eps^2}\ \,
\ffi_0(A_++itB_+,B_+,\eps^2,a_++\eta_+t,\eta_+,x)\ \,\phi_n(x),
$$
where the values of $A_+$, $B_+$, $a_+$, $\eta_+=k^*$ are those
given by (\ref{ident}), as in Theorem \ref{mai}.
The amplitude for making the transition
from level $j$ to level $n$ is given by
\bea
{\cal A}_{nj}^{(m)}(\eps)
&=&e^{-i\theta_j(\zeta)}\,e^{-\alpha(E^*)/\eps^2}\
\sqrt{\frac{B_+}{B_-}}\ \frac{e^{-i(\kappa(E^*)-k^*a_+)/\eps^2}}
{2^{m/2}\,(m!)^{1/2}}\,\left(\frac{\overline{B_-}}{B_-}\right)^{m/2}\,
H_m\left(\frac{k^*-\eta_-}{\eps\,|B_-|}\right)\nonumber\\[4mm]
&=&e^{-i\theta_j(\zeta)}\ \frac{e^{-\alpha(E^*)/\eps^2}}{\eps^m}\
\sqrt{\frac{B_+}{B_-}}\ 
\frac{e^{-i(\kappa(E^*)-k^*a_+)/\eps^2}}
{(m!)^{1/2}}\left(\frac{\sqrt{2}(k^*-\eta_-)}{B_-}\right)^m(1\,+\,O(\eps)) .
\nonumber
\eea
In particular,
$B_+$ is again given by (\ref{b+}) and the pre-exponential factor 
is of order $\eps^{-m}$. 
\end{thm}

\section{Technicalities}\label{technical}
\setcounter{equation}{0}

\noindent
{\bf Proof of lemma \ref{asesx}:}\quad
We consider only the limit $x\ra\infty$ and the choice $\sigma=+$.
The other cases are similar. In this proof, $c_n$ denotes a
finite constant that 
depends only on $n$, but may vary from line to line.

Explicitly, for any $n\in\N$,
$$
\frac{\partial^n\phantom{x}}{\partial E^n}\ \sqrt{2(E-e_j(x))}\
=\ c_n\ (2(E-e_j(x)))^{1/2-n},
$$
uniformly for $E\in\Delta$. So, the first assertion is true. Moreover,
\bea
&&\frac{\partial^n\phantom{x}}{\partial E^n}\ \left(\sqrt{2(E-e_j(x))}-
\sqrt{2(E-e_j(\infty))}\right)\quad =\nonumber\\ \nonumber
&&\qquad\qquad\qquad\quad\quad \quad\quad\quad\quad  \quad
c_n\left((2(E-e_j(x)))^{1/2-n}-(2(E-e_j(\infty)))^{1/2-n}\right).
\eea
For $n=0$, we have by (\ref{elecdecay1}),
\bea
\sqrt{2(E-e_j(x))}\,-\,\sqrt{2(E-e_j(\infty))}&=&\frac{2\,(e_j(\infty)-e_j(x))}
{\sqrt{2(E-e_j(x))}+\sqrt{2(E-e_j(\infty))}}\nonumber\\[3mm]\label{73}
&=&O(e_j(\infty)-e_j(x))\ =\ O(<x>^{-(2+\nu)}).
\eea
For $n>0$, we can write
\bea\nonumber
&&(2(E-e_j(x)))^{1/2-n}-(2(E-e_j(\infty)))^{1/2-n}\\[3mm] &=&
\frac{(2(E-e_j(x)))^{1/2}-(2(E-e_j(\infty)))^{1/2}}{(2(E-e_j(x)))^{n}}
\nonumber\\[3mm]\label{74}
&&+\quad\left(\sum_{k=0}^{n-1}\frac{(2(E-e_j(\infty)))^{1/2}}
{(2(E-e_j(x)))^{k+1}(2(E-e_j(\infty)))^{n-k}}\right)\ (2(e_j(x)-e_j(\infty)),
\eea
to which the estimate (\ref{73}) applies. The second assertion follows.

By definition,
$$
r_j^+(+,x,\eps)\ =\
-\ \int_x^\infty\
\left(\,(2(E-e_j(y)))^{1/2}-(2(E-e_j(\infty)))^{1/2}\,\right)\ dy,
$$
so that (\ref{74}) implies the estimates on
$\frac{\partial^n}{\partial E^n}r_j^+(+,x,E)$.

We now study the properties of the $c_j^\tau$'s. Again, we shall consider
$x\ra +\infty$; the other case is similar. We first compute
\bea
 a^{\tau \sigma}_{j l}(x,E)&=&-\,\frac12\,
 \frac{1}{\sqrt{k_j(x,E)k_l(x,E)}}\left(
\bra\phi_j(x),\phi'_l(x)\ket(k_j(x,E)+\tau \sigma k_l(x,E))
\phantom{ \frac{\partial}{\partial x}}\right.\nonumber\\
&& \left.\quad \quad \quad \quad
\quad\quad \quad \quad\quad \quad \quad\quad +\,
\left(\sigma\tau-\frac{k_j}{k_l}\right)\,
\frac{\bra\phi_j(x),\phi_l(x)\ket}2\,
\frac{\partial}{\partial x}k_l(x,E)\right)\nonumber\\
&=&-\,\frac12\,\frac{1}{\sqrt{k_j(x,E)k_l(x,E)}}\left(
\bra\phi_j(x),\phi'_l(x)\ket(k_j(x,E)+\tau \sigma k_l(x,E))
\phantom{ \frac{\partial}{\partial x}}\right.\nonumber\\ \label{compa}
&& \left.\quad \quad \quad \quad \quad
  \quad\quad
\quad \quad\quad\quad \quad +\,
\left(\sigma\tau-\frac{k_j}{k_l}\right)\,
\frac{\bra\phi_j(x),\phi_l(x)\ket}{2\,k_l(x,E)} e_l'(x)\right).
\eea
The presence of the factors $\bra\phi_j(x),\,\phi'_l(x)\ket$ and $e_l'(x)$,
which are independent of $E$ and decay as $1/<x>^{2+\nu}$,
implies together with
(\ref{consder}) that
\be\label{dera}
\frac{\partial^n\phantom{i}}{\partial E^n}\,a^{\tau \sigma}_{j l}(x,E)\
=\ O(<x>^{-(2+\nu)}).
\ee
We denote the coefficients $c_j^\tau$ collectively by
$$
{\bf c}(x,E,\eps)\ =\ \pmatrix{{\bf c}^+(x,E,\eps)\cr
{\bf c}^-(x,E,\eps)}\ \in\ \C^{2m},
$$
and the generator of equation (\ref{coefD}) by the $2m\times 2m$ block matrix
\bea
&& {\bf M}(x,E,\eps)=\nonumber\\[3mm] \nonumber
&&\quad \quad
\pmatrix{\vspace{2mm}
e^{i\,\int_0^x(k_j(y,E)-k_l(y,E))dy/\eps^2}\ a^{++}_{jl}(x,E)&
e^{i\,\int_0^x(k_j(y,E)+k_l(y,E))dy/\eps^2}\ a^{+-}_{jl}(x,E)\cr
e^{i\,\int_0^x(-k_j(y,E)-k_l(y,E))dy/\eps^2}\ a^{-+}_{jl}(x,E)&
e^{i\,\int_0^x(-k_j(y,E)+k_l(y,E))dy/\eps^2}\ a^{--}_{jl}(x,E)},
\eea
so that (\ref{coefD}) can be rewritten as
$$
 \frac{\partial}{\partial x}\,{\bf c}(x,E,\eps)
 \ =\ {\bf M}(x,E,\eps)\ {\bf c}(x,E,\eps).
$$
Expressing the solutions as Dyson series, we obtain
\bea\label{dyson}
&& {\bf c}(x,E,\eps)\ =\ \sum_{n=0}^\infty \int_0^x \int_0^{x_1}\cdots
 \int_0^{x_{n-1}}\\[3mm]
&& \quad \quad \quad \quad \quad \quad \times \
{\bf M}(x_1,E,\eps){\bf M}(x_2,E,\eps)\cdots {\bf M}(x_n,E,\eps)
 dx_1dx_2\cdots dx_n \ {\bf c}(0,E,\eps), \nonumber
\eea
where, because of (\ref{dera}),
$\int_0^\infty \|{\bf M}(y,E,\eps)\|dy<\infty$, uniformly
for $E\in\Delta$, we get the usual bound
$$
\|\,{\bf c}(x,E,\eps)\,\|\
\leq\ e^{\int_0^\infty\,\|{\bf M}(y,E,\eps)\|\,dy}\
\|\,{\bf c}(0,E,\eps)\,\|.
$$
By showing that $\|c(x,E,\eps)-c(y,E,\eps)\|$ is arbitrarily small
for large $x$ and $y$, we see that
$ \lim_{x\ra\infty}{\bf c}(x,E,\eps)={\bf c}(\infty,E,\eps)$ exists.
Because of the presence of the phases\\
$e^{i\int_0^x(\tau k_j(y,E)+\sigma k_l(y,E))dy/\eps^2}$ in
${\bf M}(x,E,\eps)$, whose derivatives with respect to $E$ satisfy
$$
\frac{\partial^n}{\partial E^n}\,e^{i\int_0^x(\tau k_j(y,E)+\sigma 
k_l(y,E))dy/\eps^2}\ =\ O(<x>^n),
$$
we get for $n=0,1$,
$$
\frac{\partial^n}{\partial E^n}\,{\bf M}(x,E,\eps)\ =\ O(<x>^{-(2+\nu-n)}).
$$
Thus, by the Lebesgue Dominated Convergence Theorem, we get from
(\ref{dyson}) that, as $x\ra\infty$ and uniformly for $E\in\Delta$,
$$
 \frac{\partial^n}{\partial E^n}\,{\bf c}(x,E,\eps)\ =\ O(1),
\qquad\mbox{for}\quad n=0,\,1.
$$

Finally, we consider
$$
{\bf c}(x,E,\eps)\,-\,{\bf c}(\infty,E,\eps)\ =\
-\,\int_x^\infty\,{\bf M}(y,E,\eps)\,{\bf c}(y,E,\eps)\,dy.
$$
In this expression, we use
the above properties of ${\bf M}$, ${\bf c}$, and their derivatives
to obtain the last two statements of the lemma.
\hfill\ep\\

\vskip 4mm \noindent
{\bf Proof of Lemma \ref{L2}:}\quad \\
We assume $t\neq 0$ and
rewrite the exponential factors in (\ref{sol}) as
\be\label{ippok}
e^{-i(\int_0^x\,k_j^\sigma(y,E)\,dy\,+\,tE)/\eps^2}
\ =\ 
i\,\eps^2\
\frac{\frac{\partial}{\partial E}\,
      e^{-i(\int_0^x\,k_j^\sigma(y,E)\,dy\,+\,tE)/\eps^2}}
{\left(\,t\,+\,\int_0^x\,
\frac{\partial}{\partial E}\,k_j^\sigma(y,E)\,dy\,\right)}.
\ee
Then, for each integral in (\ref{sol}), we have
\bea\nonumber
&&\hspace{-1.3cm}
\int_{\Delta}\,\frac{Q(E,\eps)\,c_j^\sigma(x,E,\eps)}{\sqrt{k_j(x,E)}}\ 
e^{-i(\int_0^x\,k_j^\sigma(y,E)\,dy\,+\,tE)/\eps^2}\ dE\\[2mm]
&&\hspace{-1cm}=\quad
\left.
i\,\eps^2\,e^{-i(\int_0^x\,k_j^\sigma(y,E)\,dy\,+\,tE)/\eps^2}\ 
\frac{Q(E,\eps)\ c_j^\sigma(x,E,\eps)}
{\sqrt{k_j(x,E)}\,\left(\,t\,+\,
   \int_0^x\,\frac{\partial}{\partial E}\,k_j^\sigma(y,E)\,dy\,\right)}
\,\right|_{E_1}^{E_2}\nonumber\\[2mm]
&&\hspace{-1cm} 
-\quad \int_{\Delta}\,\frac{\partial}{\partial E}\,
\left\{\,
\frac{Q(E,\eps)\ c_j^\sigma(x,E,\eps)}
{\sqrt{k_j(x,E)}\,\left(\,t\,+\,
   \int_0^x\,\frac{\partial}{\partial E}\,k_j^\sigma(y,E)\,dy\,\right)}
\,\right\}
\ i\,\eps^2\,e^{-i(\int_0^x\,k_j^\sigma(y,E)\,dy\,+\,tE)/\eps^2}\ dE.
\nonumber\\&&\label{woof}
\eea
The quantities $c_j^\sigma(x,E,\eps)$, $\sqrt{k_j(x,E)}$,
and their derivatives with respect to $E$
are uniformly bounded in $x$ and $E$. Also,
\bea\nonumber
\int_0^x\,\frac{\partial k_j^\sigma(y,E)}{\partial E}\,dy
&=&x\ \frac{\partial k_j^\sigma(\pm\infty,E)}{\partial E}\ +\ O(1)\\[2mm]
&=&\frac{\sigma\ x}{k_j(\pm\infty,E)}\ +\ O(1),\nonumber
\eea
uniformly for $E\in\Delta$ as $x\ra \pm\infty$.

Thus, the boundary terms in (\ref{woof}) satisfy
\bea
&&\left. i\,\eps^2\,
e^{-i(\int_0^xk_j^\sigma(y,E)dy+tE)\eps^2}\
\frac{Q(E,\eps)\,c_j^\sigma(x,E,\eps)}
{\sqrt{k_j(x,E)}\,
\left(t+\int_0^x\frac{\partial}{\partial E}k_j^\sigma(y,E)dy\right)}
\,\right|_{E_1}^{E_2}\nonumber\\[3mm]
%&&\quad\quad\quad\quad\quad\quad\quad\quad \quad\quad\quad \quad
%\quad\quad\quad
&=&O\left(\,
\frac{1}{\left|k_j^\sigma(\pm\infty,E_1)t+x+O(1)\right|}\ +\
\frac{1}{\left|k_j^\sigma(\pm\infty,E_2)t+x+O(1)\right|}
\,\right).\label{howl2}
\eea

We now apply the restrictions on $x/t$ in the statement of the Lemma.
For any choice of $j$ and $\sigma$, 
they ensure that the denominators on the right hand side of (\ref{howl2})
can be estimated, uniformly in $E$ and for large $|x|$, by
\bea
\left|\,k_j^\sigma(\pm\infty,E)t\,+\,x\,+\,O(1)\,\right|
&=&|x|\,\left|\,1\,+\,k_j^\sigma(\pm\infty,E)t/x\,+\,O(1/x)\right|
\nonumber\\[2mm]\nonumber
&\geq&|x|\,(\alpha\,+\,O(1/|x|)),
\eea
where $\alpha$ is the number that appears in the statement of the lemma.
From this, we see that the boundary terms in (\ref{woof}) are $O(1/|x|)$.

\vskip 2mm
We estimate the integral term in (\ref{woof})
in a similar way.
Under the restrictions on $x/t$ in the lemma, we obtain
\bea
&&\int_{\Delta}\,i\,\eps^2\ e^{-i(\int_0^x\,k_j^\sigma(y,E)\,dy\,+tE)/\eps^2}
\,\left(\,\left\{\,
\frac
{\frac{\partial}{\partial E}\,
   (Q(E,\eps)\,(k_j^\sigma(x,E))^{-1/2}\,c_j^\sigma(x,E,\eps))}
{t\,+\,\int_0^x\,\frac{\partial}{\partial E}\,k_j^\sigma(y,E)\,dy}\,
\right\}\,\right.\nonumber\\[2mm]
&&\qquad\qquad\qquad\quad
\left.-\quad
\frac{Q(E,\eps)\,(k_j^\sigma(x,E))^{-1/2}\,
  c_j^\sigma(x,E,\eps)\,
  \int_0^x\,\frac{\partial^2}{\partial E^2}\,k_j^\sigma(y,E)\,dy}
{\left(\,t\,+\,
  \int_0^x\,\frac{\partial}{\partial E}\,k_j^\sigma(y,E)\,dy\,\right)^2}
\right)\ dE\nonumber\\[4mm]\nonumber
&&\qquad =\quad O(1/|x|).
\eea
This implies the lemma for $t\neq 0$. When $t=0$, the estimate
(\ref{howl2}) with $t=0$ yields the result in a more direct way. 
\hfill $\ep$\\

{\bf Proof of Proposition \ref{scat}: } We can write
\bea\label{asdif}
&&\hspace{-15mm}\psi(x,t,\eps)\,-\,\psi(x,t,\eps,\pm)\ =\
\sum_{j=1,2 \ \sigma=\pm}\ \phi_j(x) \times \\
&&\hspace{-11mm}\times\
\left\{\,\int_{\Delta}\,
\frac{Q(E,\eps)\ dE}{\sqrt{2k_j(\pm\infty, E)}}\,e^{-itE/\eps^2}\,
(c_j^\sigma(x,E,\eps)-c_j^\sigma(\pm\infty,E,\eps))\,
e^{-i\int_0^xk_j^\sigma(y,E)dy /\eps^2}\nonumber\right.\\
&&\hspace{-3mm}+\ \int_{\Delta}\,
\frac{Q(E,\eps)\ dE}{\sqrt{2k_j(\pm\infty, E)}}
\,e^{-itE/\eps^2}\,c_j^\sigma(\pm\infty,E,\eps)
\nonumber\\
&&\quad\quad\quad\quad\quad\quad\quad\quad\quad\quad\quad\quad\quad
\times\quad \left(
e^{-i\int_0^xk_j^\sigma(y,E)dy /\eps^2}-
e^{-i(xk_j^\sigma(\pm\infty,E)+\omega_j^\sigma(\pm\infty,E)) /\eps^2}\right)
\nonumber\\
&&\hspace{-3mm}\left. + \int_{\Delta}
\frac{Q(E,\eps)c_j^\sigma(x,E,\eps)dE}{\sqrt{k_j(\pm\infty, E)k_j(x, E)}}
\,e^{-itE/\eps^2}
\frac{k_j(\pm\infty, E)-k_j(x, E)}{\sqrt{2k_j(\pm\infty, E)}+\sqrt{2k_j(x, E)}}
\,e^{-i\int_0^xk_j^\sigma(y,E)dy /\eps^2}\nonumber\right\}.
\eea
The first step of the proof consists of
integrating by parts to get a factor of $1/t$ according to
\bea
\int_{\Delta}\,f(x,E,\eps)\,e^{-itE/\eps^2}\,dE
&=&\left.\frac{i\,\eps^2}{t}\
f(x,E,\eps)\  e^{-itE/\eps^2}\,\right|_{E_1}^{E_2}\nonumber\\[3mm]\label{snort}
&&-\quad\frac{i\,\eps^2}{t}\
\int_{\Delta}\ \frac{\partial\phantom{i}}{\partial E}\,f(x,E,\eps)\
e^{-itE/\eps^2}\,dE.
\eea
We then bound the $L^2(\R_x)$ norm of each term that arises from
these integrations by parts, with bounds that are
uniform in $t$.

From the estimates in Lemma \ref{asesx}, we see that all the boundary terms in
(\ref{snort}) coming from (\ref{asdif}) are of order
$<x>^{-(1+\nu)}$. Thus, their $L^2$ norms are bounded, uniformly in $t$.
The integral terms in (\ref{snort}) coming from (\ref{asdif}) all have the form
\be\label{intppxt}
\int_{\Delta}\,g_j(x,E,\eps)\ e^{-i(\int_0^xk_j^\sigma(y,E)dy+tE)/\eps^2}\,
dE, \qquad j=1,2,3,
\ee
where the first integral from (\ref{asdif}) contains the function
\bea\label{g1}
g_1(x,E,\eps)&=&\left\{\,
\frac{\partial\phantom{i}}{\partial E}\,\left(\,
\frac{Q(E,\eps)}{\sqrt{k_j(\pm\infty,E)}}\
(c_j^\sigma(x,E,\eps)-c_j^\sigma(\pm\infty,E,\eps))\,
\right)\,\right\}\\[2mm] \nonumber
&-&i\,\frac{Q(E,\eps)}{\sqrt{k_j(\pm\infty,E)}}\ 
(c_j^\sigma(x,E,\eps)-c_j^\sigma(\pm\infty,E,\eps))\,\int_0^x
\frac{\partial\phantom{i}}{\partial E}k_j^\sigma(y,E)\,dy.
\eea
With the notation of Lemma \ref{asesx}, the
second integral in (\ref{asdif}) contains the function
\bea\label{g2}
g_2(x,E,\eps)&=&\frac{\partial\phantom{i}}{\partial E}\,
\left(\frac{Q(E,\eps)}{\sqrt{k_j(\pm\infty,E)}}\
c_j^\sigma(\pm\infty,E,\eps)\,\left( 1-
e^{i(r_j^\sigma(\pm,x,E)) /\eps^2} \right)\right)\\
&-&i\ \frac{Q(E,\eps)}{\sqrt{k_j(\pm\infty,E)}}\
c_j^\sigma(\pm\infty,E,\eps)\,\left( 1-
e^{i(r_j^\sigma(\pm,x,E)) /\eps^2} \right)\,\int_0^x
\frac{\partial\phantom{i}}{\partial E}k_j^\sigma(y,E)\,dy.\nonumber
\eea
The third integral contains
\bea\label{g3}\hspace{-5mm}
g_3(x,E,\eps)&=&\frac{\partial\phantom{i}}{\partial E}
\left(\frac{Q(E,\eps)\,c_j^\sigma(x,E,\eps)}{\sqrt{k_j(\pm\infty, E)k_j(x,E)}}\
\frac{k_j(\pm\infty, E)-k_j(x,E)}{\sqrt{2k_j(\pm\infty,E)}+\sqrt{2k_j(x,E)}}
\right)\\
&-&i\ \frac{Q(E,\eps)c_j^\sigma(x,E,\eps)}{\sqrt{k_j(\pm\infty, E)k_j(x, E)}}\
\frac{k_j(\pm\infty, E)-k_j(x, E)}{\sqrt{2k_j(\pm\infty, E)}+\sqrt{2k_j(x, E)}}
\,\int_0^x
\frac{\partial\phantom{i}}{\partial E}k_j^\sigma(y,E)\,dy.\nonumber
\eea

By Lemma \ref{asesx}, and the condition $\nu >1/2$, each of these 
functions $g_j(x,E,\eps)$ satisfies the following bound, uniformly in $E$,
$$
g_j(x,E,\eps)\ =\ O(<x>^{-\nu})\ \in\ L^2(\R).
$$

Therefore, we can estimate the $L^2$ norm of the corresponding expression
(\ref{intppxt}) by
$$
\int_{\R}\ \,\left|\,\int_{\Delta}\,g_j(x,E,\eps)\ 
e^{-i(\int_0^xk_j^\sigma(y,E)dy+tE)/\eps^2}\,dE\ \right|^2\ dx\
\leq\ C_1(\eps),
$$
where $C_1(\eps)$ is a finite constant that is
independent of $t$. This finishes the proof.\hfill $\ep$\\

\noindent
{\bf Proof of Proposition \ref{smaller}:} Since the argument is virtually
identical to the one presented in \cite{jp3} and \cite{joye}, we will be
rather sketchy and mainly point out the effects of the parameter $\delta$
and of the non self-adjointness of the generator $F(x,\delta)$.\\

Expressing the projector $P(x,\delta)$ as a integral of the resolvent
$(F(x,\delta)-z)^{-1}$ along a loop $L$ (or a finite number of such loops)
around the set $\sigma_1(x,\delta)$ by means of the Riesz formula,
\be\label{rf}
P(x,\delta)\ =\ -\,\frac{1}{2\pi i}\ \oint_L\,(F(x,\delta)-z)^{-1}\,dz,
\ee
we get a bound, uniform in $\delta>0$ and $ x\in \rho_\alpha$,
$$
\|P(x,\delta)\|\ \leq\ c.
$$
Indeed, for each $x\in\rho_\alpha$, we can choose the path
$L$ uniformly in $\delta$ by hypothesis.
The existence of the limits $F(\pm\infty,\delta)$
allows us actually to consider only a finite
number of distinct loops a finite distance $g/4$ away from spectrum
of $F(x,\delta)$, for all
$(x,\delta)$ . Also, uniformly in $\delta>0$,
\be\label{esres}
\|\,(F(x,\delta)-z)^{-1}\,\|\ \leq\ c,
\ee
for $z$ on the corresponding loop $L$, since
$|\det(F(x,\delta)-z)|\geq (g/4)^n$ and
$F(x,\delta)$ is uniformly bounded.
By a similar argument, using  hypothesis {\bf H4}, we get, uniformly in $\delta$
$$
\|P(x,\delta)-P(\pm\infty ,\delta)\|\ \leq\ \frac{c}{ <x>^{2+\nu}},
$$
as $x\ra\pm\infty $ in $\rho_\alpha$.
With the notation $\phantom{i}'$ for $\frac{\partial}{\partial x}$, we
get from (\ref{rf}),
$$
P'(x,\delta)\ =\
\frac{1}{2\pi i}\ \oint_L\
(F(x,\delta)-z)^{-1}\,F'(x,\delta)\,(F(x,\delta)-z)^{-1}\,dz.
$$
Thus, hypothesis {\bf H4} yields, uniformly in $\delta$,
$$
\|\,P'(x,\delta)\,\|\ \leq\ \frac{c}{<x>^{2+\nu}},
$$
and a similar uniform estimate for $K(x,\delta)=[P'(x,\delta),P(x,\delta)]$,
\be\label{esk}
\|\,K(x,\delta)\,\|\ \leq\ \frac{c}{<x>^{2+\nu}}.
\ee
The operator $K$ is the generator of the intertwining operator $W$
defined by
$$
W'(x,x_0,\delta)\ =\ K(x,\delta)\,W(x,x_0,\delta),
\qquad\mbox{with}\qquad W(x_0,x_0,\delta)\ =\ \un.
$$
It satisfies 
\be\label{inter}
W(x,x_0,\delta)\,P(x_0,\delta)\ =\ P(x,\delta)\,W(x,x_0,\delta),
\ee
for all $(x,\delta)$ (including $x=\pm\infty$).

Following \cite{jp5}, we construct a hierarchy of generators.
Let $F_0(x,\delta)=F(x,\delta)$,\\ $P_0(x,\delta)=P(x,\delta)$, and
$K_0(x,\delta)=K(x,\delta)$.
For $q\in\N^*$, we inductively define
$$
F_q(x,\delta,\eps)\ =\ F(x,\delta)\,-\,\eps^2\,K_{q-1}(x,\delta,\eps),
$$
assuming $\eps$ is small enough so that the spectrum of $F_q$ is separated into
two disjoint parts corresponding to those of $F$.
We define $P_q(x,\delta,\eps)$ to be
the spectral projector for $F_q$ corresponding to $P(x,\delta)$ as
$\eps\ra 0$ by perturbation theory. Then,
$$
K_q(x,\delta,\eps)\ =\ [P_q'(x,\delta,\eps),\,P_q(x,\delta,\eps)].
$$
Sections II.A and II.B of \cite{jp5} and (\ref{esres}) and (\ref{esk})
show that there exist
constants $\eps^*>0$, $r>0$, $\Gamma>0$ and $c>0$, all independent of $\delta>0$,
such that for all $\eps<\eps^*$, all $x\in\R$, and $q=q^*=[r/\eps^2]$,
\bea
&&\label{smk} \| K_{q^*-1}(x,\delta,\eps)\|\ \leq\ \frac{c}{<x>^{2+\nu}},\\
&&\| K_{q^*}(x,\delta,\eps)-K_{q^*-1}(x,\delta,\eps)\|\ \leq\ 
c\ \frac{e^{-\Gamma/\eps^2}}{<x>^{2+\nu}}.\nonumber
\eea
We define
\bea
&&\label{pert}
F_*(x,\delta,\eps)\ =\ F_{q^*}(x,\delta,\eps)\
=\ F(x,\delta)-\eps\  K_{q^*-1}(x,\delta,\eps),\\ \nonumber
&& P_*(x,\delta,\eps)=P_{q^*}(x,\delta,\eps),\\ \nonumber
&& K_*(x,\delta,\eps)\ =\ K_{q^*}(x,\delta,\eps),
\eea
and the evolution operators $W_*$ and $\Xi_*$ by
\bea\nonumber
&&W_*'(x,x_0,\delta,\eps)\ =\ K_*(x,\delta,\eps)\ W_*(x,x_0,\delta,\eps),
\qquad\mbox{with}\qquad W(x_0,x_0,\delta,\eps)\ =\ \un,\\[3mm]
\hspace{-2.5cm}\mbox{and}\quad&&
i\,\eps^2\,\Xi_*'(x,x_0,\delta,\eps)\ =\
W_*(x_0,x,\delta,\eps)\,F_*(x,\delta,\eps)\,W_*(x,x_0,\delta,\eps)\,
\Xi_*(x,x_0,\delta,\eps), \nonumber\\[3mm]
&& \hskip 7cm \mbox{with}\qquad\Xi_*(x_0,x_0,\delta,\eps)\ =\ \un.\label{phi}
\eea
The intertwining property (\ref{inter}) still holds with the $*$ indices.
Therefore, $\Xi_*$ satisfies
$$
[\Xi_*(x,x_0,\delta,\eps),\,P_*(x_0,\delta,\eps)]\ \equiv\ 0,
\qquad\mbox{for all}\qquad x\in\R.
$$
It follows from the definitions that the operator
$$
V_*(x,x_0,\delta,\eps)\ =\ W_*(x,x_0,\delta,\eps)\ \Xi_*(x,x_0,\delta,\eps)
$$
satisfies
$$
i\,\eps^2\,V_*'(x,x_0,\delta,\eps)\ =\
(F_*(x,\delta,\eps)\,+\,i\,\eps^2\,K_*(x,\delta,\eps))\ V_*(x,x_0,\delta,\eps)
$$
and
\be\label{inter*}
V_*(x,x_0,\delta,\eps)\ P_*(x_0,\delta,\eps)\ =\
P_*(x,\delta,\eps)\ V_*(x,x_0,\delta,\eps).
\ee
Moreover,
\bea\label{752}
&&U_\eps(x,x_0,\delta)\,-\,V_*(x,x_0,\delta,\eps)\ =\\[3mm]
&& \hskip 2cm i\,\int_{x_0}^x\,V_*(x_0,y,\delta,\eps)\,
(K_{q^*}(y,\delta,\eps)-K_{q^*-1}(y,\delta,\eps))\,
U_\eps(y,x_0,\delta)\,dy.\nonumber
\eea

The proposition will follow from
\be\label{753}
U_\eps(x,x_0,\delta)\,-\,V_*(x,x_0,\delta,\eps)\ =\ O(e^{-\Gamma/\eps^2}),
\ee
(\ref{inter*}) and
$$
\lim_{x\ra\pm\infty}\ P_*(x,\delta,\eps)\ =\ P(\pm\infty,\delta),
$$
due to (\ref{smk}) and (\ref{pert}).

To prove (\ref{753}), we first prove 
that $V_*$ is uniformly bounded in $x, x_0$, and $\eps$.
The analysis leading
to Lemma \ref{asesx} implies that $U_\eps$ is uniformly bounded in
$x, x_0$, and $\eps$.
This property is a consequence on the fact that the
eigenvalues of $F$ are simple
and real, so that  the decomposition (\ref{solad})
holds and the singular exponential factors are phases.
Note that the lack of orthogonality of the
eigenprojectors of $F(x,\delta)$ makes
the bound on $U_\eps$ dependent on $\delta$.

Choose $B(\delta)>0$, such that
$\|U_\eps(x,x_0,\delta)\|\,\le\,B(\delta)$.
From (\ref{752}) we get the inequality
$$
\|V_*(x,x_0,\delta,\eps)\|\ \le\ B(\delta)\
\left(\,1\,+\,\int_{x_0}^x\ \sup_{y_0,y}\,
\|V_*(y,y_0,\delta,\eps)\|\
\frac{C\,e^{-\Gamma/\eps^2}}{\langle\,x\,\rangle^{2+\nu}}\,\right)\,dx,
$$
for some $C$.
This implies that for some $\widetilde{C}$, the quantity
$\ds v(\eps,\delta)\ =\ \sup_{x_0,x}\,\|V_*(x,x_0,\delta,\eps)\|$ satisfies
$$
v(\eps,\delta)\ \le\ B(\delta)\
\left(\,1\,+\,\widetilde{C}\,v(\eps,\delta)\,e^{-\Gamma/\eps^2}\,\right).
$$
This implies
$$v(\eps,\delta)\ \le\
\frac{B(\delta)}{1\,-\,\widetilde{C}\,B(\delta)\,e^{-\Gamma/\eps^2}}
\ \le\ \widetilde{v}(\delta),
$$
where $\widetilde{v}(\delta)$
is uniformly bounded for sufficiently small $\eps$.

We now use (\ref{752}) again to see that
\bea\nonumber
\|\,U_\eps(x,x_0,\delta)\,-\,V_*(x,x_0,\delta,\eps)\,\|&\le&
\widetilde{v}(\delta)\ \,
\int_{\r}\ 
\frac{B(\delta)\,C\,e^{-\Gamma/\eps^2}}{\langle\,x\,\rangle^{2+\nu}}
\ dx\\[3mm] \nonumber
&\le&\widetilde{C}_1\ e^{-\Gamma/\eps^2}.
\eea
This proves (\ref{753}) and completes the proof of the proposition.
\hfill\ep\\

{\bf Proof of Lemma \ref{dec}: }
Degenerate perturbation theory for self-adjoint matrices and hypothesis
{\bf H5} (see \cite{j}) show that there exist
$f(z,\delta)$ and $\rho(z,\delta)$, analytic in $z$ for fixed $\delta$,
and $C^1$ as functions of $(z,\delta)$, such that
\bea\label{eive}
e_j(z,\delta)&=&f(z,\delta)\ -\ \frac12\ \sqrt{\rho(z,\delta)}\\
e_n(z,\delta)&=&f(z,\delta)\ +\ \frac12\ \sqrt{\rho(z,\delta)}.\nonumber
\eea
where, as $(z,\delta)\ra (0,0)$,
$$
f(z,\delta)\ =\ f(0,0)\,+\,O(|z|+\delta)\ =\ e_c\,+\,O(|z|+\delta),\qquad
\mbox{and}\qquad\rho(z,\delta)\ =\ O(|z|^2+\delta^2).
$$
Moreover, $\rho(z,\delta)$ has two simple zeros, the complex crossing points,
$z_0(\delta)$ and $\bar{z}_0(\delta)$ that have $z_0(\delta)=O(\delta)$.
For concreteness, we arbitrarily choose $e_j<e_n$ on the real axis,
although this is irrelevant for the analysis.
Thus, by {\bf H7}, we can write
$$
\sqrt{2\,(E-e_j(z,\delta))}\ =\
\sqrt{2\,(E-f(z,\delta))}\ \left(\,1\,+\,\frac{\sqrt{\rho(z,\delta)}}
{2\,(E-f(z,\delta))}\,\right)^{1/2},
$$
where $(E-f(z,\delta))$ and its inverse are analytic in $\rho_\alpha$,
uniformly in $E\in \Delta$. Moreover,
$$
\left(\,1+\frac{\sqrt{\rho(z,\delta)}}
{2\,(E-f(z,\delta))}\,\right)^{1/2}
\ =\ 1\ +\ \frac12\
\frac{\sqrt{\rho(z,\delta)}}{2\,(E-e_c)}\ +\ O(|z|^2+\delta^2).
$$
Therefore, since $\sqrt{2\,(E-f(z,\delta))}$ is analytic, and we can choose
the loop $\zeta$ encircling $z_0(\delta)$ or $\bar{z}_0(\delta)$ to
satisfy $|\zeta|=O(\delta)$,
we see that
$$
\int_\zeta\ \sqrt{2(E-e_j(z,\delta))}\ dz\quad =\quad
\frac{1}{\sqrt{2(E-e_c)}}\
\int_\zeta\ \frac{\sqrt{\rho(z,\delta)}}{2}\ dz\ +\ O(\delta^3),
$$
and 
$$
 \int_\zeta\ \frac{\sqrt{\rho(z,\delta)}}{2}\ dz\ \,=\,\ O(\delta^2).
$$
In these two expressions,
$\ds\int_\zeta\,\frac{\sqrt{\rho(z,\delta)}}{2}\,dz\
=\ \int_\zeta\,e_j(z,\delta)\,dz$
due to the analyticity of $f$ in (\ref{eive}). Taking the
imaginary part yields the first statement of the lemma.
Note that we do not have
to worry about sign issues because Theorem \ref{PERCO} ensures the decay rate,
$\ds\Im\int_\zeta\,\sqrt{2(E-e_j(z,\delta))}\,dz$, is positive.\\

The two other statements follow from similar considerations for the integrals
\bea\nonumber
\frac{\partial\phantom{x}}{\partial E}\
\Im\int_\zeta\,\sqrt{2\,(E-e_j(z,\delta))}\ dz&=&
\Im\int_\zeta\,\frac{1}{\sqrt{2\,(E-e_j(z,\delta))}}\ dz
\qquad\qquad\mbox{and}\\[2mm] \nonumber
\frac{\partial^2\phantom{x}}{\partial E^2}\
\Im\int_\zeta\,\sqrt{2\,(E-e_j(z,\delta))}\ dz&=&
-\ \Im\int_\zeta\,\frac{1}{\left(2\,(E-e_j(z,\delta))\right)^{3/2}}\ dz.
\eea
\hfill\ep\\

\noindent
{\bf Proof of Lemma \ref{icet}:}

Consider first the  minimization of the negative of the real part of 
the exponent.

Since $\gamma_j(E)$ tends to zero with $\delta$
(absent in the notation), if $\delta$ is small enough, we must look
for minima in a neighborhood of $E_0$ that satisfy the equation
$$
\alpha'(E)\ =\ g\,(E-E_0)\,+\,\Im \gamma_j'(E)\,+\,O(E-E_0)^2\ =\ 0.
$$
We consider the absolute minimum $E^*$ of $\alpha$ and assume
it is unique.
By Lemma \ref{dec}, $\Im \gamma_j'(E)<0$, so $E^*>E_0$.
Note that $E^*$ does not depend on $x$ or $t$. Also,
$\Im \gamma_j^{(n)}(E)=o(\delta)$, $n=0,1,2$,
uniformly in $E$. So, we can assume $E^*$ is non-degenerate since
$$
\alpha''(E^*)\ =\ g\,+\,\Im \gamma_j''(E^*)\,+\,O(E^*-E_0)\ >\ 0.
$$

In terms of the variable $k\in [k_1,k_2]$, we view
$T$ as the (scaled) inverse Fourier transform of the function
$$
R(k,t,\eps)\ =\ \sqrt{2\pi\eps^2}\ e^{-\alpha(E(k))/\eps^2}\
\tilde{P}(E(k),\eps)\ \sqrt{k}\
e^{-i\kappa(E(k))/\eps^2}\,e^{-it(k^2/2+e(\infty))/\eps^2}\,
\chi_{[k_1,k_2]}(k),
$$
where $\chi_{S}(\cdot)$ is the characteristic function of the set
$S$.  That is
$$
T(x,t,\eps)\ =\ ({\cal F}_\eps^{-1}\,R(\cdot,t,\eps))(x), 
$$
where ${\cal F_\eps}$ is defined
by
$$
({\cal F_\eps}\,g)(x)\ =\
\frac{1}{\sqrt{2\pi\eps^2}}\ \int_{\r}\ g(k)\,e^{-ikx/\eps^2}\,dk.
$$

With the variable $k\in [k_1,k_2]$ we have
$$\left.
\frac{\partial^2\phantom{i}}{\partial k^2}\,
\alpha(E(k))\,\right|_{k^*}\ =\ {k^*}^2\ \alpha''(E^*),
$$
and expanding around $k^*$, 
\bea\hspace{-5mm}
T(\eps, x, t)&=&e^{-\alpha(E^*)/\eps^2}\nonumber\\
&&\times\quad\int_{[k_1,k_2]}
\sqrt{k}\ 
\tilde{P}(E(k),\eps)\
e^{-\frac{\frac{\partial^2}{\partial k^2}\alpha(E(k))|_{k^*}}{2\eps^2}(k-k^*)^2}
\,e^{O((k-k^*)^3)/\eps^2}\, 
e^{-i\beta(k,x,t)/\eps^2}\,dk,\nonumber
\eea
where the negative of the imaginary part of the exponent is denoted by
$$
\beta(k,x,t)\ =\ t\,\left(\frac{k^2}{2}+e_2(\infty)\right)\,+\,
\kappa(E(k))\,-\,x\,k.
$$

We now introduce $\mu(\eps)=\eps^s>0$, with $2/3<s<1$. 
It goes to zero in such a way that
$$
\mu(\eps)/\eps >> 1 \ \ \  \mbox{and} \ \ \ \mu(\eps)^3/\eps^2 <<1 .
$$
Because $E^*$ is a unique absolute minimum, the behavior of $\alpha(E)$ close to
$E^*$, and the assumption (\ref{condd}) on $P$, we can reduce the
integration range in $T$ to
$[k^*-\mu(\eps),\,k^*+\mu(\eps)]$ at the expense of a relative
error whose $L^2$ norm is of order $O(\eps^\infty)$, 
uniformly $t$. More precisely, 
$$
T(x,t,\eps)\ =\ (({\cal F}_\eps^{-1}(R_1+R_2))(\cdot,t,\eps))(x), 
$$
where
\bea\nonumber
R_1(k,t,\eps)&=&\chi_{[k^*-\mu(\eps), k^*+\mu(\eps)]}(k)\ R(k,t,\eps),
\qquad\qquad\mbox{and}\\ \nonumber
R_2(k,t,\eps)&=&\chi_{[k^*-\mu(\eps), k^*+\mu(\eps)]^{C}}(k)\ R(k,t,\eps).
\eea
For some $a^*>0$ and $r>0$,
$$
|R_2(k,t,\eps)|\ \leq\ r\  e^{-\alpha(E^*)/\eps^2}\ e^{-a^* (\mu(\eps)/\eps)^2}
\ \eps\ |\sqrt{k}\,\tilde{P}(E(k),\eps)|.
$$
Hence, by the Parseval identity, uniformly in $t$, we have
$$
\|{\cal F}_\eps^{-1}(R_2)(\cdot,t,\eps))\|\ =\
\left\{\,\int_{[k^*-\mu(\eps), k^*+\mu(\eps)]^{C}}\,|R_2(k,t,\eps)|^2
\,dk\,\right\}^{1/2}
\ =\ O(e^{-\alpha(E^*)/\eps^2}\eps^\infty).
$$
In the remaining integral containing $R_1$, we further estimate
\bea\label{780}
&&e^{O(k-k^*)^3/\eps^2}\ =\ 1+O(\mu(\eps)^3/\eps^2)\ =\ 1+O(\eps^{3s-2}),
\eea
and
\bea\nonumber
&&\sqrt{k}\ \tilde{P}(E(k),\eps)\ =\ 
\sqrt{k^*}\ \tilde{P}(E^*,\eps)\,+\,O(\mu(\eps))\ =\
\sqrt{k^*}\ {P}(E^*,\eps)\,+\,O(\eps^{s}+\eps^2).
\eea
The contribution of order $\eps^2$ comes from the error in the
computation of the coefficient $c_n^-$.
Using the Parseval identity again with uniform bounds on the exponential
factors of $R_1$, we see that the contribution to $T$ coming from the
error term
$O(\eps^{s})$ is bounded uniformly in $t$ in the $L^2(\R_x)$ norm
by  $O(e^{-\alpha(E^*)/\eps^2}\eps^{1+2s})$. Similarly, the error term 
stemming from (\ref{780}) yields an error in the
$L^2(\R_x)$ norm of order $O(e^{-\alpha(E^*)/\eps^2}\eps^{4s-1})$.

To compute the leading term, we expand $\beta(\cdot, x,t)$ around $k^*$ as
\bea
\beta(k,x,t)&=&t\,E^*\,+\,\kappa(E^*)\,-\,x\,k^*\nonumber\\
&+&(k-k^*)\,\left(\,k^*\,t\,+
\left.\frac{\partial}{\partial k}\kappa(E(k))\right|_{k^*}
-\,x\,\right)\nonumber\\
&+&\frac{(k-k^*)^2}{2}\ \left(\,t\,+\,
\left.\frac{\partial^2}{\partial k^2}\kappa(E(k))\right|_{k^*}\,\right)
\nonumber\\
&+&\frac{(k-k^*)^3}{6}\ 
\left.\frac{\partial^3}{\partial k^3}\kappa(E(k))\right|_{\tilde{k}},
\label{782}
\eea
where
$\tilde{k}$ lies between $k$ and $k^*$, and
the third derivative is independent of $t$ and $x$. The last
term in (\ref{782}) gives rise to a contribution which is of order 
$O(e^{-\alpha(E^*)/\eps^2}\eps^{4s-1})$ in the $L^2(\R_x)$ norm,
uniformly in $t$, as above.

Therefore, in the $L^2$ sense,
\bea
T(\eps, x, t)&=&e^{-\alpha(E^*)/\eps^2}\ 
e^{-i(tE^*+\kappa(E^*)-xk^*)/\eps^2}
\nonumber\\
&&\times\quad\left\{\,
\int_{[k^*-\mu(\eps),k^*+\mu(\eps)]}
\sqrt{k^*}\,P(E^*,\eps)\,
e^{-i\frac{(k-k^*)}{\eps^2}
(k^*t+\frac{\partial }{\partial k}\kappa(E(k))|_{k^*})-x)
}\right.\nonumber\\[2mm]
&&\qquad\quad\left.
\times\ 
e^{-\frac{(k-k^*)^2}{2\eps^2}(\frac{\partial^2 }{\partial k^2}\alpha(E(k))|_{k^*}+
i(t+\frac{\partial^2 }{\partial k^2}\kappa(E(k))|_{k^*}))}\,dk\,+\,O(\eps^p)\, 
+\, O(\eps^\infty)
\right\},\nonumber
\eea
where $p=\min(1+2s,4s-1)\in(0,\,3)$ can be chosen arbitrarily close to 3.
Again, at the cost of an error whose $L^2$ norm is
$O(e^{-\alpha(E^*)/\eps^2}\eps^\infty)$,
uniformly in $t$, we can extend the
interval of integration to the whole real line
and compute the Gaussian
integral explicitly according to the formula (for $\Re M >0$)
$$
\int_{-\infty}^\infty \sqrt{k^*}\,e^{-(M(k-k^*)^2/2+iN(k-k^*))/\eps^2}\,
dk\ =\
\frac{\eps}{\sqrt{k^*}}\
e^{-\frac{N^2}{2\eps^2 M}}\
\left(\sqrt{\frac{2\pi}{M}}\,(k^*-iN/M)\right).
$$
We then get the result with
\bea\nonumber
M&=&\frac{\partial^2 }{\partial k^2}\alpha(E(k))|_{k^*}
\,+\,i\,\left(t+\frac{\partial^2 }{\partial k^2}\kappa(E(k))|_{k^*}\right),
\qquad\mbox{and}\\ \nonumber
N&=&k^*t\,+\,\frac{\partial }{\partial k}\kappa(E(k))|_{k^*}\,-\,x.
\eea\hfill\ep\\

{\bf Proof of Lemma \ref{outgaus}: } The first assertion is
straightforward. The second follows from the identity
\bea\hspace{-8mm}
\ffi_0(A_+(t), B_+(t),\eps^2, a_+(t), \eta_+(t), x)\ x&=&
\ffi_0(A_+(t), B_+(t),\eps^2, a_+(t), \eta_+(t), x)\ (x-a_+(t))
\nonumber\\[2mm]
&&+\quad \ffi_0(A_+(t), B_+(t),a_+(t),\eps^2, \eta_+(t), x,)\ a_+(t).
\nonumber\eea
The first term is $O({\eps})$ in $L^2(\R)$ by scaling, and the
second is of order $a_+(t)=k^*t(1+O(1/|t|))$ for $|t|$ large. We insert
this in the first part of the lemma to obtain the second part
as $t\ra\pm\infty$.\\ \phantom{DONE!!!!}\hfill \ep\\

\end{document}